\documentclass[10pt, aps, groupedaddress, twocolumn, nofootinbib, prd, tightenlines]{revtex4-1}
\usepackage{latexsym}
\usepackage{amsmath}
\usepackage{amsfonts}
\usepackage{amsbsy}
\usepackage{amssymb}
\usepackage{epsfig}
\usepackage{mathrsfs}
\usepackage{epsfig}
\usepackage{bbm}
\usepackage{xcolor}

\newtheorem{defin}{Definition}

\newtheorem{lem}{Lemma}

\newtheorem{theorem}{Theorem}

\newtheorem{cor}{Corollary}

\usepackage{hyperref}
\hypersetup{
    colorlinks,%
    citecolor=blue,%
    filecolor=blue,%
    linkcolor=blue,%
    urlcolor=blue
}

\begin{document}

\title{{\bf Properties of the Hamiltonian Renormalisation\\
and its application to quantum mechanics on the circle}}

\author{Benjamin Bahr}
\email[]{benjamin.bahr@desy.de}
\affiliation{II. Institute for Theoretical Physics, University of Hamburg\\ Luruper Chaussee 149, 22761 Hamburg, Germany}
\author{Klaus Liegener}
\email[]{klaus.liegener@desy.de}
\affiliation{II. Institute for Theoretical Physics, University of Hamburg\\ Luruper Chaussee 149, 22761 Hamburg, Germany}

\date{\today}

\begin{abstract}
We consider the Hamiltonian renormalisation group flow of discretised one-dimensional physical theories. In particular, we investigate the influence the choice of different embedding maps has on the RG flow and the resulting continuum limit, and show in which sense they are, and in which sense they are not equivalent as physical theories. We are furthermore elucidating the interplay of the RG flow and the algebras operators satisfy, both on the discrete and the continuum. Further, we propose preferred renormalisation prescriptions for operator algebras guaranteeing to arrive at preferred algebraic relations in the continuum, if suitable extension properties are assumed. Finally, we introduce a weaker form of distributional equivalence, and show how unitarily inequivalent continuum limits, which arise due to a choice of different embedding maps, can still be weakly equivalent in that sense.

\end{abstract}


\maketitle

\section{Introduction and motivation}

Upon constructing a quantum theory of a given system one is often faced with many ambiguities. A common way to attack those is by starting to demand the behaviour of the quantum system at a certain (coarse) resolution. Afterwards, the theory at other scales can be determined by implementing suitable compatibility criteria, e.g. cylindrical consistency.

This method is known as {\it renormalisation} \cite{GL:54,WK:74} in the context of the covariant path integral quantisation and let to many prominent applications \cite{BB:01,GRS:14}. There are many formulations of renormalisation, but the philosophy employed throughout this paper comes closest to the block spin transformations of lattice gauge theories \cite{Kad:66,Kogut:79,Cre:84}. While numerical investigations have proven succesful under various approximations, there are still open questions remaining on the conceptual side, e.g. the chocie of how to relate coarse degrees of freedom with those on finer scales (to which we continue to refeer to as {\it embedding map}). These issues become paramount when there is no comparison to experiments yet, e.g. if turning towards avenues for quantum gravity \cite{Nagy:14,Loll:19,Bahr:2009qc,Dittrich:2014ala,Bahr:2014qza,Ste:20}.

On the Hamiltonian side, utilising the renormalisation group (RG) for constructing quantum field theories (QFT) is best enunciated in the language of inductive limits \cite{KR86,Jan88,Sau98,Thi07,Lang:2017beo}. As of today the Hamiltonian renormalisation is less developed then its covariant counterpart and thus many conceptual questions remain unanswered as well (such as the role of embedding maps, the final interpretation of the limit Hilbert space etc.). In this paper, we address some of these problems and demonstrate the consequences exemplary for the case of 1-particle quantum mechanics on the circle.\\

In section \ref{section:2_General} we will discuss the properties of the Hamiltonian RG on the general level, using the formulation  of inductive limits:
an inductive family is a family of Hilbert spaces endowed with suitable embedding maps. From the physical point of view, we may interpret each Hilbert space as the collection of those states which can be fully described at some coarse resolution $M$. 

Here the resolution $M$ serves as a generalisation of a UV-cutoff scale, in the sense that it specifies, up to which resolution information of the system is accessible. In QFT, this is the equivalent of e.g.~lattice spacing, while in quantum mechanics with finitely many degrees of freedom, this resolution specifies a finite-dimensional subspace of the full Hilbert space. The advantage of this notion of resolution and coarse graining lies in the fact that the mathematical framework does not necessarily require a background metric, which allows for a notion of background-independent renormalisation scheme, which is specifically useful for approaches to quantum gravity, see e.g.~\cite{Dittrich:2011zh, Bahr:2011aa, Bahr:2012qj,  Dittrich:2012jq, Dittrich:2014ala, Bahr:2014qza, Dittrich:2016tys, Bahr:2016hwc, Bahr:2017klw, Lang:2017beo, Lang:2017yxi, Lang:2017oed, Lang:2017xrb,Liegener:2020dbc,Thiemann:2020cuq,Morinelli:2020uea}.

The fact that there exists also finer resolutions $M'$ by which those states can be described without losing information is encapsulated via the embedding maps which embed the Hilbert space of resolution $M$ into the one of resolution $M'$. For an inductive family the embedding maps are suitably chosen, such that they finally allow the restoration of a inductive limit or {\it continuum} Hilbert space, i.e. the collection of states at all scales.

The choice of embedding map encapsulates significant physical information, and can make the renormalisation of the system harder or easier, depending on whether it fits together well with the dynamics of the system. Indeed, the central point of approaches like MERA or TNR (see e.g.~ \cite{levin,guwen,vidal-evenbly}) is to construct the correct embedding maps for a given Hamiltonian. In this case, the embedding maps contain all the information about the continuum vacuum state.

Assuming that such an inductive family of Hilbert spaces is given, we focus our attention on observables of the continuum Hilbert space. We recall in section \ref{section:2-1_IndLim} that a family of operators on the Hilbert spaces of finite resolution which obeys weak (/strong) cylindrical consistency can be promoted to a bilinear form (/operator) on the inductive limit Hilbert space. A typical application of the RG is constructing such a cylindrical consistent family starting from some initial ad-hoc choice, i.e. a discretisation.

Every fixed point of the RG flow yields a cylindrical consistent family, however the precise form of the family is dependent on both the choice of embedding map as well as the initial discretisation. However, when being interested in the whole set of possible fixed points, we show in section \ref{section:2-3_UniEqu} that actually one of those data is redundant: one may either fix the embedding map once and for all and study the RG flow of all possible initial discretisations (or vice versa) without loosing any cylindrical consistent families.

In general, one is not interested in a cylindrically consistent family for a single observable, but usually a whole algebra thereof (constituting the set of possible questions to ask to the system). Such an algebra is characterised by the algebraic relations between its elements (e.g. commutator relations) and we investigate in section \ref{section:2-4_AlgRel} how the RG flow of algebraic relations between operators at discrete level translates to operators on the continuum Hilbert space (hereby we often need to make strong assumptions that the bilinear forms are extendible). In subsection \ref{section:2-5_ModRG} we reformulate our findings in a constructive criterion to obtain cyl. cons. algebras obeying given algebraic relations.

Finally, in section \ref{section:2-6_DistEmb} we ask about the interpretation of observables in the continuum Hilbert space as part of the space of distributions $\mathcal{D}'$. Closable operators can be understood as suitable restrictions of maps from $A':\mathcal{D}'\to \mathcal{D}'$. Simultaneously, different inequivalent fixed point theories may be embeddable into $\mathcal{D}'$ with the help of suitable {\it faithful embeddings}. In case that these embeddings approximate different restrictions of the same map $A'$, we call them {\it weakly equivalent}.

In order to show-case our findings, section \ref{section:3_Example} considers discretised quantum mechanics on the circle. We study two embedding maps and their RG flow starting from a common initial discretisation leading to two inequivalent fixed points. However, we show that ultimately both fixed points are weakly equivalent as both approximate the continuum.

In section \ref{section:4_Conclusion} we conclude with an outlook for further research directions.

\section{General Properties of the Hamiltonian RG}
\label{section:2_General}

\subsection{Inductive limits}
\label{section:2-1_IndLim}

\noindent We recall basic facts from the theory of inductive limits and streamline the notation. We omit the corresponding proofs and refer to \cite{KR86} or the appendix of \cite{Lang:2017beo} for further details.\\

The starting point of each inductive family are partial Hilbert spaces $\mathcal{H}_M$, which we take to be finite-dimensional (if not specified otherwise). The label $M$ belongs to a partially ordered, directed set $\mathcal{I}$, i.e.~there is a relation $M\leq M'$, such that for every two $M,M'$ there is a $M''$ such that $M\leq M''$ and $M'\leq M''$. We define {\it embedding maps} for $M\leq M'$ as
\begin{eqnarray}
I_{M\to M'}\;:\;\mathcal{H}_M\:\longrightarrow\;\mathcal{H}_{M'}
\end{eqnarray}

\noindent which are isometries and satisfy
\begin{eqnarray}
I_{M'\to M''}I_{M\to M'}\;=\;I_{M\to M''},
\end{eqnarray}

\noindent as well as $I_{M\to M}=\text{id}_{\mathcal{H}_M}$. Given this data, we can define
\begin{eqnarray}
\mathcal{D}_\infty\;:=\;\bigsqcup_M\mathcal{H}_M\big/\sim
\end{eqnarray}

\noindent where $\psi\sim I_{M\to M'}\phi$ for $\phi\in\mathcal{H}_M$ and $\psi\in\mathcal{H}_{M'}$. The completion of the pre-Hilbert space $\mathcal{D}_\infty$ is the \emph{continuum Hilbert space}
\begin{eqnarray}
\label{continuumHS_closure}
\mathcal{H}_\infty\;:=\;\overline{\mathcal{D}_\infty}.
\end{eqnarray}

\noindent Continuum embedding maps $I_M:\mathcal{H}_M\to \mathcal{H}_\infty$ are then given by
\begin{eqnarray}
I_M\psi\;:=\;[\psi].
\end{eqnarray}

\noindent  Then, if $M\leq M'$ it is clear that
\begin{eqnarray}
I_M=I_{M'}I_{M\to M'} 
\end{eqnarray}
\\

On Hilbert spaces, we denote operators by $\hat{A}$. Each operator $\hat{A}$ determines a bilinear\footnote{Technically, the form is sesquilinear and not bilinear, but to comply with standard terminology in this field, we also use the term bilinear form.} form

\begin{eqnarray}
A(\psi,\phi)\;:=\;\langle \psi,\,\hat{A}\phi\rangle.
\end{eqnarray}

\noindent Note the opposite is not true: Assume that $A:\mathcal{D}\times\mathcal{D}\to \mathbb{C}$, is a bilinear form, where $\mathcal{D}\subset\mathcal{H}$ is any dense domain containing an orthonormal basis $\{e_n\}_n$. Then there is an operator $\hat{A}:\mathcal{D}\to \mathcal{H}$ with $A(\psi,\phi)=\langle \psi,\hat{A}\phi\rangle$ if and only if
\begin{eqnarray}\label{Eq:ConditionBilinearFormIsAnOperator}
\sum_n|A(e_n,\psi)|^2\;<\;\infty
\end{eqnarray}
\noindent for all $\psi\in\mathcal{D}$. We also use the notation
\begin{eqnarray}
A(m,n)\;:=\;A(e_m,e_n)
\end{eqnarray}

\noindent for the kernel with respect to an ONB $\{e_n\}$. This means that iff $\sum_n|A(m,n)|^2<\infty$ for all $m$, then $\hat{A}$ can be defined on at least the finite linear span of the $e_n$.\\

In the following we consider constructing bilinear forms and operators on $\mathcal{H}_\infty$, if those on $\mathcal{H}_M$ are given.

\begin{defin}
A family of bilinear forms $\{\hat A_M\}_M$ on each $\mathcal{H}_M$ is said to satisfy \emph{weak cylindrical consistency} if for every $M\leq M'$ one has 
\begin{eqnarray}\label{weak_cyl_consistency}
A_{M'}(I_{M\to M'}\psi,\,I_{M\to M'}\phi)\;=\;A_M(\psi,\,\phi).
\end{eqnarray}

\noindent $\forall \psi,\phi\in\mathcal{H}_M$. Given a cylindrically consistent family of bilinear forms, we can define a bilinear form on the continuum
\begin{eqnarray}
A_\infty(\psi,\phi)\;:\;\mathcal{D}_\infty\times\mathcal{D}_\infty\;\longrightarrow\;\mathbb{C}
\end{eqnarray}

\noindent in a straightforward way.\\
Some families allow directly to define operators on the continuum level: we say that a family of operators $\{\hat{A}_M\}_M$ is \emph{strongly cylindrically consistent}, if for $M\leq M'$
\begin{eqnarray}\label{strong_cyl_consistency}
\hat{A}_{M'}I_{M\to M'}\;=\;I_{M\to M'}\hat{A}_M.
\end{eqnarray}
\end{defin}

Note that the bilinear forms defined by strongly cylindrically consistent operators are weakly cylindrically consistent, but not the other way round. We expect interesting physics to be encoded in operators on $\mathcal{H}_\infty$ which are the extension of weakly consistent forms but not satisfying (\ref{strong_cyl_consistency}). This is due to the fact that (\ref{strong_cyl_consistency}) implies $\hat A_M$ to preserve subspace of resolution $M$, which is typically violated in applications.\\

We make some further assumption about the label set $\mathcal{I}$, which are not strictly necessary, but which make our life much easier, and comply with most interesting physical situations:\\
\noindent Firstly, we assume that there exists a smallest element $M_0\in\mathcal{I}$. This will correspond to the maximally possible coarse graining, and often appears in situations in which no infra-red divergences appear, e.g.~with compact spatial manifolds.\\
\noindent Secondly, we assume that there are \emph{amenable subsequences}, i.e.~sequences $\mathcal{I}_0\subset\mathcal{I}$  such that for every $M\in \mathcal{I}$ there is an $M'\in\mathcal{I}_0$ with $M\leq M'$.\\
\noindent If both assumptions are true, one can replace $\mathcal{I}$ by $\mathcal{I}_0\simeq \mathbb{N}$, and obtain equivalent Hilbert spaces, operators, etc. In particular, it will make the RG flow equations simpler, which is why we will assume we can work with the label set
\begin{eqnarray}\label{Eq:AmenableLabelSet}
\mathcal{I}\;=\;\{1,2,4,\ldots, 2^L,\ldots \,|\,L\in\mathbb{N}\}
\end{eqnarray}

\noindent from now on, without loss of generality.

\subsection{The RG flow equations}
\label{section:RG-flow}

The RG flow is a tool to construct (weakly) cylindrically consistent families of bilinear forms\footnote{Indeed, given a family of partial Hilbert spaces $\mathcal{H}_M$ and maps $I_{M\to M'}$ which are not necessarily isometries, one can use the RG flow to modify the inner products on each $\mathcal{H}_M$ in order to arrive at a cylindrically consistent family of Hilbert spaces of which one can take the continuum limit as described above. See \cite{Lang:2017beo} for details. In this article however, we always assume that the maps $I_{M\to M'}$ are isometries.}.

In practice, one starts with a sequence of bilinear forms ${}^{(o)}A_M$, which are not necessarily cylindrically consistent. Then, we define
\begin{eqnarray}\label{RG_limit_operator}
A_M(\psi,\phi)\;:=\; \lim_{M'\to\infty} \!\! {}^{(o)}A_{M'}(I_{M\to M'}\psi,\, I_{M\to M'}\phi)\;\;\;\;\;\;
\end{eqnarray}
\noindent where $\psi,\phi\in\mathcal{H}_M$. Technically, one can define the flow iteratively\footnote{The flow is defined generally, the iterative version exists only if the label set has an amenable subsequence.}
\begin{eqnarray}
{}^{(n+1)}A_M(\psi,\phi)\;:=\;{}^{(n)}A_M(I_{M\to 2M}\psi,  I_{M\to 2M}\phi)\;\;\;\;
\end{eqnarray}

\noindent and consider the limit of the ${}^{(n)}A_M$ for large $n$. If this limit exists, the resulting $A_M$ define a cylindrically consistent family of bilinear forms. 

Note, that for the RG flow of bilinear forms, it is never guaranteed that the limit converges to a meaningful form a priori. This is highly dependent on the choice of initial discretisation and embedding map. A counter-example is presented in the paragraph on the momentum operator in \ref{section:3_D-DiscOfOper}, where an initial discretisation together with certain choice of embedding map gives a good limit, while another choice of embedding map sends the same initial discretisation to zero.

\subsection{Unitary Equivalence of Embedding maps}
\label{section:2-3_UniEqu}

Under quite general assumptions, different discretisations are equivalent on the level of Hilbert spaces.

To be precise, we assume that we have given two families of partial Hilbert spaces $\mathcal{H}^{(i)}_M$, as well as embedding maps $I^{(i)}_{M\to M'}$, $i=1,2$ based on the same label set $\mathcal{I}$, which we assume to be (\ref{Eq:AmenableLabelSet}). The respective continuum Hilbert spaces be $\mathcal{H}_\infty^{(i)}$. 

\begin{lem}
Let $\mathcal{H}^{(i)}_{M=1}\cong \mathbb{C}$ for $i=1,2$. Also, assume that  $\dim(\mathcal{H}^{(1)}_M)=\dim(\mathcal{H}^{(2)}_M) < \infty$ for all $M$. \\
Then, there exists a unitary isomorphism 
\begin{align}
\xi_\infty : \mathcal{H}^{(1)}_\infty \longmapsto \mathcal{H}^{(2)}_\infty 
\end{align}
\end{lem}

{\it Proof.} First, we show that for all $M$ there exist bijections $\xi_M: \mathcal{H}^{(1)}_M \to \mathcal{H}^{(2)}_M$ such that $\xi_M I_{M\to M'}^{(1)} = I_{M\to M'}^{(2)} \xi_{M'}$ for $M\leq M'$.

We build $\xi_{M}$ by recursion over $M=2^L$, starting from $\xi_{M=1}=1$. For $2M=2^{L+1}$ we define the unitary isomorphism on the image of embedding maps as:
\begin{align*}
\xi_{2M}\; :\; {\rm Im}&(I^{(1)}_{M\to 2M}) \to {\rm Im} (I^{(2)}_{M\to 2M})\subset \mathcal{H}^{(2)}_M\\[5pt]
\psi&\;\;\longmapsto I_{M\to 2M}^{(2)} \xi_{M} (I_{M\to 2M}^{(1)})^{-1} \psi
\end{align*}
\noindent which is well-defined because $I_{M\to 2M}^{(1)}$ is invertible on its image. Now, consider the orthogonal complement $V^{(i)}_M\perp {\rm Im}(I^{(i)}_{M\to 2M})$. We have $\dim(V^{(1)}_M)=\dim(V^{(2)}_M)$ and thus $V^{(1)}_M\cong V^{(2)}_M$. To be precise, we can choose some orthonormal bases $\{e^{(i)}_\ell\}_\ell$ of both $V^{(i)}_M$ and finish the construction of $\xi_{2M}$ by defining:
\begin{align}
\langle e^{(2)}_{\ell}, \xi_{2M} e^{(1)}_{\ell'} \rangle := \delta_{\ell,\ell'}
\end{align} 

\noindent This gives a sequence of $\xi_M$ obviously satisfying
\begin{eqnarray}
\xi_M I_{M\to M'}^{(1)} \;=\; I_{M\to M'}^{(2)} \xi_{M'}\qquad \text{for }M<M'.
\end{eqnarray}

\noindent Thus, this defines an isometry $\xi_\infty:\mathcal{D}_\infty^{(1)}\to\mathcal{D}_\infty^{(2)}$, which can be completed to the respective continuum Hilbert spaces $\mathcal{H}_\infty^{(i)}$.\hfill $\square$\\

This finishes the proof for the finite-dimensional case. In case of infinite-dimensional separable $\mathcal{H}_M$, the proof runs similarly, with one additional condition.

\begin{lem}
Let $I^{(i)}_{M\to M'}$ be isometries on $\mathcal{H}^{(i)}_M$ and for all $M$ assume $\dim(\mathcal{H}^{(i)}_M)=\dim((I_{M\to2M}\mathcal{H}^{(i)}_M)^\perp)=\infty$, but being separable.\\
Then, as all infinite-dimensional, separable Hilbert spaces are isomorphic to $\ell^2$ by using Zorn's Lemma, there also exists an isomorphism $\xi$ on the inductive limit Hilbert spaces.
\end{lem}

A corollary of those lemmas is that each cylindrical consistent family of operators on $\mathcal{H}^{(1)}_M$ can be unitarily mapped to a cylindrical consistent family on $\mathcal{H}^{(2)}_M$ such that the inductive limit operator families are equivalent as well. Further, also the algebraic relation between the operators in the continuum will be the same.\\
That is, the resulting Hilbert spaces and operators in the continuum are unitarily equivalent and thus the precise choice of embedding map does not matter at least conceptually. Hence, while in principle one has to choose an initial discretisation and an embedding map in order to study some RG flow, conceptually it suffices to restrict the attention to one initial discretisation and several different embedding maps (as will be done in section \ref{section:3_Example}) or vice versa.\\
However, it must be noted that in practice, one typically needs to do some approximations while computing an involved RG flow. Now, different embedding maps will more or less susceptible to different approximation procedures, thus the map should be chosen such that possible errors during the approximations process do not influence the resulting physics. For further discussion on this subject see \cite{Dittrich:2013xwa} and references therein.

\subsection{Algebraic relations of operators}
\label{section:2-4_AlgRel}

Physical observables are encoded as operators on the continuum Hilbert space. The RG flow constructs bilinear forms, which can be turned into operators if criterion (\ref{Eq:ConditionBilinearFormIsAnOperator}) is satisfied. However, whether the correct physics is implemented, is governed by the algebra satisfied by the respective operators. 

Here, one encounters an interesting tension between the continuum operators $\hat{A}$ and the partial operators $\hat{A}_M$ defined at a specific coarse graining scale. Namely, the partial and continuum operators satisfy the same algebra if the partial operators are strongly cylindrically consistent, i.e.~satisfy (\ref{strong_cyl_consistency}). However, in practice the operators on the continuum Hilbert space are described by only weakly cylindrically consistent operators, i.e.~whose bilinear forms satisfy (\ref{weak_cyl_consistency}), and for which the correct algebra does not have to hold both on $\mathcal{H}_\infty$ and $\mathcal{H}_M$. I.e.~while the continuum physics is correctly represented, at finite discretisation scale there will be anomalies. These can be interpreted as discretisation artefacts. An interesting example will be given in section \ref{section:3_D-DiscOfOper}.\\
Nevertheless, given partial operators $\hat{A}_M$, $\hat {B}_M,\ldots$ for which the limit (\ref{RG_limit_operator}) exists, one can make certain statements about the RG flow of their algebraic relations.

\begin{theorem}\label{Theorem1}
Let $\{\hat{A}_M,\hat{B}_M\}_M$ be two weakly cylindrical consistent families of operators, whose inductive limit quadratic forms $A_\infty,B_\infty$ satisfy (\ref{Eq:ConditionBilinearFormIsAnOperator}) on $\mathcal{D}_\infty$, i.e.~exist as densely defined closable operators $\hat{A}_\infty,\hat{B}_\infty$. Also assume that these leave $\mathcal{D}_\infty$ invariant.

Then, the RG flow of ${}^{(o)}\hat{C}_M:= \hat{A}_M\hat{B}_M$ -- defined via (\ref{RG_limit_operator}) -- converges, and gives rise to a quadratic form $C_\infty$ on $\mathcal{D}_\infty$ which exists as operator $\hat{C}_\infty$ on $\mathcal{D}_\infty$ satisfying
\begin{eqnarray}\label{Eq:ProductOfOperators}
\hat{C}_\infty\;=\;\hat{A}_\infty\; \hat{B}_\infty.
\end{eqnarray}
\end{theorem}

{\it Proof.} See appendix \ref{proof_thm}. The proof is indeed given for an arbitrary monomial.\hfill $\square$\\

\noindent There is an interesting consequence of this theorem, which alludes to our earlier statement: Even though the $\hat{A}_M$, $\hat{B}_M$ are cylindrically consistent, the ${}^{(o)}\hat{C}_M=\hat{A}_M\hat{B}_M$ are not. However, as a starting point for the RG flow, they flow to $\hat{C}_M$ which are the partial operators of $\hat{C}_\infty=\hat{A}_\infty \hat{B}_\infty$. If the $\hat{A}_M$, $\hat{B}_M$ are not strongly consistent, however, the RG flow is nontrivial, and even though (\ref{Eq:ProductOfOperators}) holds, one has 
\begin{eqnarray}
\hat{C}_M\;\neq\;\hat{A}_M\hat{B}_M.
\end{eqnarray} 

\noindent This impacts the operator algebras defined on the continuum Hilbert space, since the above statement extends to arbitrary monomials, and commutators. In particular, even though the correct algebra might be satisfied at the continuum level, at the level of the partial Hilbert spaces, anomalies can arise.\\
Thus, one should not check whether a set of weakly consistent operators satisfies certain algebraic relations, but could instead test whether the algebraic relations are satisfied in the limit of the RG flow in order to guarantee the correct continuum physics.\\

It should be noted that this feature is conceptually related, but different from the breaking of symmetries by discretisations, discussed in \cite{Bahr:2009ku, Bahr:2009qc}. In particular, the $\hat{A}_M$, $\hat{B}_M$ and $\hat{C}_M$ are the perfect discretisations from the continuum, but their algebra is still anomalous due to discretisation arefacts.

\subsection{Modified RG flow}
\label{section:2-5_ModRG}

Typically, one is interested in the quantisation of some specific algebra of observables $\mathcal{A}$. Can one use the RG flow to enforce its algebraic relations for the continuum operators?

By virtue of its definition, every algebra is endowed with a map $\mathcal{A}\times \mathcal{A} \to \mathcal{A}$, capturing its defining algebraic relations. As example, consider for the bilinear product "$\cdot$" and assume that for all $A,A'\in\mathcal{A}$ there exist finitely many $z_a\in\mathbb{C}$ such that
\begin{eqnarray}\label{algebraic_relations}
A \cdot A' = \sum_{\alpha=1}^N z_\alpha A(\alpha)\in\mathcal{A}
\end{eqnarray}
where $\alpha$ is some labelling of the elements in $\mathcal{A}$.\\
Given the set-up of section \ref{section:RG-flow}, one might attempt to implement a quantisation of $\mathcal{A}$ in the following way: first, take some initial choice of operators ${}^{(o)}\hat A_M$ on $\mathcal{H}_M$. For each, construct the weak cylindrical consistent families $\hat A_M$ following (\ref{RG_limit_operator}) and collect those to obtain the candidate-quantisation set $\mathcal{A}_M$. Even under the assumption that the bilinear forms on $\mathcal{D}_\infty$ turn out to be extendible to operators, one must finally check, whether those operators fulfil the algebraic relations (\ref{algebraic_relations}) of $\mathcal{A}$.\\
In general, nothing guarantees that the algebra closes. However, if the resulting bilinear forms are extendible to operators, let us note that a sufficient criterion for obtaining the correct algebra comes in a slight variation of theorem \ref{Theorem1}:
 if $\forall \hat A_M,\hat A'_M\in\mathcal{A}_M$ one has
\begin{eqnarray}\label{RG_limit_consistent_algebra}
\sum_{\alpha=1}^N z_\alpha \hat A_M(\alpha)=\lim_{M'\to\infty} I^\dagger_{M\to M'} (\hat A_{M'} \hat A_{M'}')  I_{M\to M'}
\end{eqnarray}
then the correct algebraic relations are obtained for the operators in the continuum.\footnote{Note, that the right-hand side of (\ref{RG_limit_consistent_algebra}) is already weakly cylindrical consistent and allows the reconstruction of bilinear forms.}\\

Ultimately, it would be advantageous to have a {\it constructive} prescription which returns operators satisfying the correct algebraic relations in the continuum. In some situations, this can be indeed achieved:

\begin{theorem}
Given an algebra $\mathcal{A}$ whose algebraic relation can be brought into the following form: $\forall A\in \mathcal{A}$ there exist $z_{\alpha,\alpha'}$ such that
\begin{eqnarray}
A = \sum_{\alpha,\alpha'=1}^N z_{\alpha,\alpha'} A(\alpha)A(\alpha')
\end{eqnarray}
Starting from some initial data ${}^{(o)}A_M$, the fixed points of the modified RG flow
\begin{align}\label{RG_flow_consistent_algebra}
{}^{(n+1)}A_M := \sum_{\alpha,\alpha'=1}^N z_{\alpha,\alpha'} \lim_{M'\to\infty} &I^\dagger_{M\to M'}\times \\
& {}^{(n)}A(\alpha){}^{(n)}A(\alpha')  I_{M\to M'}\nonumber
\end{align}
resurrect the correct algebra in the continuum if being extendible to operators on $\mathcal{D}_\infty$, i.e. satisfy (\ref{Eq:ConditionBilinearFormIsAnOperator}).
\end{theorem} 

In situations where this theorem applies, starting from some suitable initial discretisation one may study the RG flow of (\ref{RG_flow_consistent_algebra}) under a given embedding map. It will lead either (i) to a trivial fixed point, (ii) not converge, or (iii) to a theory which obeys the correct algebra for $M\to\infty$ however not necessarily for finite $M$.

\subsection{Distributional Embedding}
\label{section:2-6_DistEmb}

Finally, although not all possible fixed pointed will algebras agree in their properties, often one may interpret them as approximating the same continuum theory via distributional embeddings.

First, we introduce a suitable  space of distributions: Assume that we have a (finite-dimensional) manifold $\mathcal{M}$ with non-degenerate metric, and the space
\begin{eqnarray}\label{Eq:FunctionSpace}
\mathcal{D}\;:=\;C_0^\infty(\mathcal{M})
\end{eqnarray}

\noindent which can be given a nuclear topology, yielding the Gelfand triple
\begin{eqnarray}
\mathcal{D}\;\subset\;\mathcal{H}\;\subset\;\mathcal{D}',
\end{eqnarray}

\noindent with the Hilbert space $\mathcal{H}=L^2(\mathcal{M},d\rm{vol})$, and $\mathcal{D}'$ being the topological dual of (\ref{Eq:FunctionSpace}), the members of which can be regarded as distributions over $\mathcal{M}$. In the following, we write $\langle \varphi,\,f\rangle$ to denote the pairing of the distribution $\varphi\in \mathcal{D}'$ and the function $f\in\mathcal{D}$.\\

Next we assume that we have a collection of (finite-dimensional) Hilbert space $\mathcal{H}_M$ and embedding maps $I_{M\to 2M}$, which are used to define the continuum Hilbert space
\begin{eqnarray}
\mathcal{H}_\infty = \overline{\mathcal{D}_\infty}
\end{eqnarray}

\noindent as in (\ref{continuumHS_closure}). \emph{A priori} the two Hilbert spaces $\mathcal{H}$ and $\mathcal{H}_\infty$ do not have anything to do with one another. To connect them, we define the notion of a \emph{faithful embedding}:
\begin{defin}
A \emph{faithful embedding} of the $\mathcal{H}_M$ into $\mathcal{D}'$ is a collection of injective anti-linear maps
\begin{eqnarray}\label{Eq:FaithfulEmbedding_01}
\phi_M\;:\;\mathcal{H}_M\;\longrightarrow\;\mathcal{D}'
\end{eqnarray}

\noindent which satisfy 
\begin{eqnarray}\label{Eq:FaithfulEmbedding_02}
\phi_{2M}\,\circ\,I_{M\to 2M}\;=\;\phi_M
\end{eqnarray}
\end{defin}

By construction, a faithful embedding defines a map
\begin{eqnarray}
\phi_\infty\;:\;\mathcal{D}_\infty\;\longrightarrow\;\mathcal{D}',
\end{eqnarray}

\noindent which is a linear isomorphism onto its image. We also call this map the embedding, if no confusion arises.\\

Note that a faithful embedding satisfies that for all $\psi\in \mathcal{D}_\infty$ there is an $f_\psi\in\mathcal{D}$ such that\footnote{This is equivalent to the fact that there is an $f$ such that $\phi_\infty(\psi)\neq 0$, which follows from the injectivity of $\phi_M$ on each $\mathcal{H}_M$.}
\begin{eqnarray}\label{Eq:Separating_Dense}
\big|\langle \phi_\infty(\psi),\,f_\psi\rangle\big|\,>\,\|\psi\|.
\end{eqnarray}

In general, comparison of the two continuum theories on $\mathcal{H}$ and $\mathcal{H}_\infty$ is straightforward if $\phi_\infty$ can be extended to $\mathcal{H}_\infty$:
\begin{defin}
(1) A faithful embedding $\phi_M$ is called \emph{regular}, if it satisfies that, for every $f\in \mathcal{D}$ there exists a constant $K_f>0$ such that
\begin{eqnarray}\label{Eq:RegularFaithfulEmbedding}
\Big\vert\langle \phi_M(\psi), \,f\rangle\Big\vert\;\leq\;\Vert \psi \Vert \,K_f
\end{eqnarray}
\noindent for all $M$ and all $\psi\in\mathcal{H}_M$.\\
(2) We call a faithful, regular embedding \emph{separating}, if condition (\ref{Eq:Separating_Dense}) also holds for all $\psi\in\mathcal{H}_\infty$. 
\end{defin}
\begin{lem}
(i) A \emph{regular} embedding $\phi_\infty$ has an extension to all of $\mathcal{H}_\infty$ (also denoted by $\phi_\infty$).\\
(ii) A \emph{separating} embedding has an extension $\phi_\infty$ on $\mathcal{H}_\infty$ that is injective.
\end{lem}
To see (i), consider a Cauchy sequence $\{\psi_n\}_n$ in $\mathcal{D}_\infty$. Then, for any function $f\in \mathcal{D}$ one has, by regularity, that
\begin{eqnarray}
\Big\vert\langle \phi_\infty(\psi_n)-\phi_\infty(\psi_m), \,f\rangle\Big\vert\;\leq\;\Vert \psi_n-\psi_m \Vert \,K_f,
\end{eqnarray}

\noindent so $\langle \phi_\infty(\psi_n)-\phi_\infty(\psi_m), \,f\rangle$ is a Cauchy sequence in $\mathbb{R}$, which converges. Since convergence in $\mathcal{D}'$ is pointwise, the sequence $\{\phi_\infty(\psi_n)\}_n$ converges in $\mathcal{D}'$.\\
For (ii), let $\psi_n$ be a Cauchy sequence in $\mathcal{D}_\infty$ converging to $\psi\in\mathcal{H}_\infty$. Assume $\phi_\infty(\psi)= 0$, then
\begin{eqnarray}
0\;=\;\big|\langle\phi_\infty(\psi),f\rangle\big|\;>\;\|\psi\|
\end{eqnarray}

\noindent which implies $\psi=0$. Hence the extension is injective.\\

\begin{cor}
A faithful, regular, separating embedding lets us realise $\mathcal{H}_\infty$ as a subspace of $\mathcal{D}'$, with its own inner product. 
\end{cor}

It can happen that the thus realised $\phi_\infty(\mathcal{H}_\infty)$ and the original $\mathcal{H}$ have only the zero in common. Still, a closable operator $\hat{A}$ densely defined on $\mathcal{D}\subset \mathcal{H}$ can be carried over to a linear map
\begin{eqnarray}
\hat{A}'\;:\;\mathcal{D}'\;\longrightarrow\;\mathcal{D}'
\end{eqnarray}

\noindent via:
\begin{eqnarray}
\langle\hat{A}'\varphi,\,f\rangle\;:=\;\langle\varphi, \,\hat{A}^\dag f\rangle.
\end{eqnarray}

\noindent This way the operator $\hat{A}$ can be made into a linear map on a larger distributional space, which then in turn can be again restricted to other subspaces. This allows comparison of the continuum Hilbert space $\mathcal{H}$ and observabels thereon with operators which arise as the RG fixed point on $\mathcal{H}_\infty$.\\
However, in practical situations, the RG maps $I_{M\to M'}$ might motivate faithful embeddings $\phi_M$ which are neither regular nor separating. In fact, we will encounter such a situation later on in the example of section \ref{Sec:WeakEquivalenceOfi=1and2}. For this case, it is useful to introduce a version of \emph{weak equivalency} between discrete and continuous theories:

\begin{defin}\label{Def:approx_continuum}
Given a family of cylindrically consistent operators $\hat{A}_M$ on the $\mathcal{H}_M$. Assume that there exist:
\begin{itemize}
\item A dense subspace $\mathcal{D}_A\subset \mathcal{H}_\infty$,
\item a faithful embedding $\{\phi_M\}_M$,
\end{itemize}

\noindent such that:
\begin{itemize}
\item The bilinear form $A_\infty$ is extendible to an operator $\hat{A}_\infty:\mathcal{D}_A\to\mathcal{D}_A$,
\item The faithful embedding $\phi_\infty$ can be extended to $\phi_\infty:\mathcal{D}_\infty+\mathcal{D}_A\to \mathcal{D}'$,
\item $\phi_\infty(\mathcal{D}_A)\subset \mathcal{D}'$ is dense in the weak-*-topology.
\end{itemize}

\noindent Then, the discretised system is said to \emph{approximate the continuum} if there is a densely defined, closable  operator $\hat{A}:\mathcal{D}\to\mathcal{D}$ such that 
\begin{eqnarray}\label{Eq:OperatorsApproximateTheContinuum}
\phi_\infty\hat{A}_\infty(\phi_\infty)^{-1}{}_{\big\vert_{\phi_\infty(\mathcal{D}_A)}}\;=\;\hat{A}'{}_{\big\vert_{\phi_\infty(\mathcal{D}_A)}}
\end{eqnarray}
\end{defin}

\noindent A few remarks are in order: The above notion of weak equivalence states that there is a way, in which the partial operators $A_M$ approximate the continuum operator $\hat{A}$. However, since they live on potentially different Hilbert spaces, this approximation has to be established carefully. In particular, the continuum bilinear form $A_\infty$ might exist as an operator on $\mathcal{D}_\infty$, but might not leave it invariant. It can happen that it has to be extended to some larger space $\overline{\mathcal{D}}=\mathcal{D}_A+\mathcal{D}_\infty$, and only its restriction to $\mathcal{D}_A$ is a closable operator with invariant domain of definition.\footnote{The momentum operator for $i=2$ is an example for this, see \ref{Sec:MomentumOperators}.} On the other hand, in order to make contact with distributions, the faithful embedding $\phi_\infty$ needs to be extendible to $\overline{\mathcal{D}}$ as well. Therefore, this notion of weak equivalence is a subtle condition on the interplay of the operators $A_M$ and the $\phi_M$. We will see an example for this in section \ref{Sec:WeakEquivalenceOfi=1and2}.

The denseness of $\phi_\infty(\mathcal{D}_A)\subset \mathcal{D}'$, together with the given denseness of $\mathcal{D}\subset \mathcal{D}'$ \cite{friedlander1998introduction}, means that states in either space can be approximated by states from the other space in the sense of distributions. From a physical point of view, this means that by performing measurements by taking inner products with elements in $\mathcal{D}$ with finite precision, there is no way of distinguishing whether a state is in either $\mathcal{D}$ or $\phi_\infty(\mathcal{D}_A)$. Of course, if one has access to the respective spectra of the operators with infinite precision, one might distinguish which system one is working with, since the systems $(\mathcal{H},\,\hat{A})$ and $(\mathcal{H}_\infty,\hat{A}_\infty)$ do not have to be unitarily equivalent. However, one can argue that that this knowledge is unattainable if one is working with only finite measurement precision.

Finally, we call two discretisations {\it weakly equivalent}, if both approximate the continuum. In fact, in that case they also approximate each other in the above sense.

\section{Discretised Quantum mechanics on the circle}
\label{section:3_Example}

\noindent This section, puts the lessons learned from the previous section to an explicit test:
Let us consider the Weyl algebra of exponentiated operators on the unit circle. In some regards, this is mathematically much more convenient than position and momentum operators for quantum mechanics on the circle.\footnote{For instance, it is well-known that even though $\hat{x}$ and $\hat{p}$ exist as operators on $\mathcal{H}:=L^2([0,1],dx)$, they do not satisfy the standard canonical commutation relations, since the largest domain on which $[\hat{x},\hat{p}]$ can be defined is $\{0\}$. } To showcase the equivalence properties, we consider inductive families with two different embedding maps $I^{(i)}_{M\to M'}$.

\subsection{Continuum theory}
\label{section:3_1_Continuum}

\noindent We begin with the original continuum Hilbert space $\mathcal{H}=L^2(S^1)\simeq L^2([0,1))$. In the following, we represent states in $\mathcal{H}$ as functions
\begin{eqnarray}
[0,1)\,\ni\,x\,\longmapsto\,f(x)\,\in\,\mathbb{C}
\end{eqnarray}

\noindent with the inner product
\begin{eqnarray}
\langle f,\,g\rangle\;=\;\int_0^1dx\;\overline{f(x)}g(x).
\end{eqnarray}

\noindent  On $\mathcal{H}$, we define operators $\hat{U}$ and $\hat{T}(l)$ with $l\in[0,1)$
\begin{eqnarray}\label{Eq:FundamentalWeylOperators}
(\hat{U}f)(x)\;&:=&\;e^{2\pi i x}f(x)\\[5pt]
(\hat{T}(l)f)(x)\;&:=&\;
\left\{
\begin{array}{cc}
f(x+l)&\qquad \text{ if }x+l<1\\[5pt]
f(x+l-1)&\qquad \text{ if }x+l\geq 1
\end{array}
\right.\nonumber
\end{eqnarray}

\noindent One can think of these as $\hat{U}=\exp(2\pi i\hat{x})$ and $\hat{T}(l)=\exp(il\hat{p}/\hbar)$. They satisfy the relations
\begin{eqnarray}\label{Eq:Weyl_Relations_Continuum}
\hat{U}\,\hat{T}(l)\,\hat{U}^{\dag}\;=\;e^{2\pi i l}\,\hat{T}(l).
\end{eqnarray}

Later on, in subsection \ref{section:3_D-DiscOfOper}, we will consider to a discretisation of the operators $\hat	U,\; \hat T (l)$, on two different families of inductive Hilbert spaces $\mathcal{H}_M^{(i)}$, which we construct in the following subsection.

\subsection{Discretisations of $\mathcal{H}$}
We introduce two different ways to discretise the continuum Hilbert space, by representing it as the inductive limit of two different families $\{\mathcal{H}^{(i)}_M, I^{(i)}_{M\to M'}\}_M$ with $i=1,2$.

For that purpose, we consider discretisations of the unit interval $[0,1]$ with finite evenly-space lattices. These are labelled by the number of their points $M=2^L$, $L=0,1,2,\ldots$. The lattice with $M$ points $m=0,1,\ldots, M-1$ contains the values of the interval
\begin{eqnarray}
x_m^{(M)}\;=\;\frac{m}{M}\;\in\;[0,1).
\end{eqnarray}

\noindent For each discretisation, we consider the Hilbert space 
\begin{eqnarray}
\mathcal{H}_M^{(1)}\;=\;\mathcal{H}_M^{(2)}\;:=\;\mathbb{C}^M.
\end{eqnarray}

\noindent Each element $f\in\mathcal{H}_M^{(i)}$ can be described by a function $f:\{0,1,\ldots, M-1\}\to\mathbb{C}$, with value $f_m$. The inner product is given by
\begin{eqnarray}
\langle f,\,g\rangle_M\;=\;\frac{1}{M}\sum_{m=0}^{M-1}\overline{f_m}g_m.
\end{eqnarray}

\noindent We define the Kronecker-delta in the $m$-th element as

\begin{eqnarray}
({}^{(M)}\delta^m)_n\;:=\;\delta ^m{}_n,\qquad n,m\in\{0,1,\ldots,M-1\},
\end{eqnarray}

\noindent as well as 
\begin{eqnarray}\label{Eq:ONB}
e^m_M\;:=\;\sqrt{M}{}^{(M)}\delta^m,
\end{eqnarray}

\noindent  which form an ONB for $\mathcal{H}_M$.  Next, we consider two different embedding maps:
\begin{eqnarray}
\begin{aligned}
I^{(1)}_{M\to 2M}\;&:\;\mathcal{H}^{(1)}_M\;\longrightarrow\;\mathcal{H}^{(1)}_{2M}\\[5pt]
I^{(1)}_{M\to 2M}\;&:\;e^m_M\;\mapsto\;e_{2M}^{2m}\label{Eq:EmbeddingMap01}
\end{aligned}
\end{eqnarray}

\noindent and

\begin{eqnarray}
\begin{aligned}
I^{(2)}_{M\to 2M}\;&:\;\mathcal{H}^{(2)}_M\;\longrightarrow\;\mathcal{H}^{(2)}_{2M}\\[5pt]\label{Eq:EmbeddingMap02}
I^{(2)}_{M\to 2M}\;&:\;e_M^m\;\mapsto\;\frac{1}{\sqrt{2}}\left(e_{2M}^{2m}\,+\,e_{2M}^{2m+1}\right).
\end{aligned}
\end{eqnarray}

\noindent From these follow arbitrary $I^{(i)}_{2^L\to 2^{L'}}$ for $L'>L$, and consequently result in an inductive limit Hilbert space
\begin{eqnarray}\label{Eq:ContinuumHilbertSpaceDefinition}
\mathcal{H}^{(i)}_\infty\;:=\;\lim_{M\leftarrow}\mathcal{H}_M^{(i)}\;=\;\overline{D^{(i)}_\infty}.
\end{eqnarray}

\noindent The dense subspace
\begin{eqnarray}
\mathcal{D}_\infty^{(i)}\;:=\;
\bigsqcup_{m}\mathcal{H}_M^{(i)}\Big/\sim
\end{eqnarray}

\noindent is endowed with the equivalence relations for either $i=1,2$ generated by
\begin{eqnarray}
f\;\sim\;I_{M\to 2M}^{(i)}f
\end{eqnarray}

\noindent for any $f\in\mathcal{M}^{(i)}$ and all $M$. We also denote the canonical unitary embedding maps
\begin{eqnarray}\label{Eq:CanonicalEmbeddingMaps}
\begin{aligned}
I^{(i)}_M\;&:\;\mathcal{H}_M^{(i)}\;\longrightarrow\;\mathcal{H}_\infty^{(i)}\\[5pt]
I^{(i)}_M\;&:\;f\;\longmapsto\;[f].
\end{aligned}
\end{eqnarray}

\subsection{Unitary equivalence}

It is clear that $\mathcal{H}_\infty^{(1)}$ and $\mathcal{H}_\infty^{(2)}$ are unitarily equivalent, since both are separable Hilbert spaces. Even further though, there is a unitary equivalence which is consistent with the two discretisations, in the sense that it intertwines the embedding maps (see subsection \ref{section:2-3_UniEqu}):

{\bf Claim:} There exists a family of unitary maps
\begin{eqnarray}
\xi_M\;:\;\mathcal{H}_M^{(1)}\;\longrightarrow\;\mathcal{H}_M^{(2)}
\end{eqnarray}

\noindent which satisfy
\begin{eqnarray}\label{Eq:IntertwiningUnitary}
\xi_{2M}I^{(1)}_{M\to 2M}\;=\;I^{(2)}_{M\to 2M}\xi_{M}.
\end{eqnarray}

\noindent {\it Proof.} To construct these unitary maps, we first note that there is a unitary isomorphism
\begin{eqnarray}\label{Eq:UsefulHilbertSpaceIsomorphism}
\chi_M\;:\;\mathcal{H}_M^{(i)}\otimes \mathcal{H}_2^{(i)}\;\longrightarrow\;\mathcal{H}^{(i)}_{2M}
\end{eqnarray}

\noindent given by
\begin{eqnarray}
\chi_M\;:\;{}^{(M)}\delta^m\otimes{}^{(2)}\delta^{m'}\;\longmapsto\;{}^{(2M)}\delta^{2m+m'}
\end{eqnarray}

\noindent for $m=1,\ldots, M-1,m'=0,1$. In what follows, we will suppress $\chi_M$, for legibility, and implicitly use the isomorphism (\ref{Eq:UsefulHilbertSpaceIsomorphism}).  This will allow to compute the unitary maps by induction over $M$. \\
First, it is clear that $\xi_1:\mathbb{C}\to\mathbb{C}$ is given by the identity map. From this we can see that (up to a complex number), the unique $\xi_2$ satisfying (\ref{Eq:IntertwiningUnitary}) for $M=1$ is given by
\begin{eqnarray}\label{Eq:UnitaryMapForMEqualsTwo}
\xi_2\;=\;\frac{1}{\sqrt{2}}\left(
\begin{array}{rr}
1\; & -1 \\ 1 \;& 1
\end{array}
\right),
\end{eqnarray}

\noindent where (\ref{Eq:UnitaryMapForMEqualsTwo}) is the matrix representation with respect to the orthonormal basis elements ${}^{(2)}\delta^m$, $m=0,1$. Furthermore, we define
\begin{eqnarray}\label{Eq:InductiveDefinitionUnitaryMaps}
\xi_{2M}\;:=\;\xi_M\otimes \xi_2
\end{eqnarray}
\noindent for $M=2^L$, using the isomorphism (\ref{Eq:UsefulHilbertSpaceIsomorphism}). Since, with respect to this isomorphism, the embedding maps can be written as
\begin{eqnarray}
\begin{aligned}
I^{(1)}_{M\to 2M}\left[{}^{(M)}\delta^m\right]\;&=\;\sqrt{2}\;{}^{(M)}\delta^m\otimes {}^{(2)}\delta^0\\[5pt]
I^{(2)}_{M\to 2M}\left[{}^{(M)}\delta^m\right]\;&=\;{}^{(M)}\delta^m\otimes\big( {}^{(2)}\delta^0+{}^{(2)}\delta^1\big)
\end{aligned}
\end{eqnarray}

\noindent This shows that (\ref{Eq:InductiveDefinitionUnitaryMaps}) defines an intertwining map, as can be checked by direct computation. For instance, the unitary maps for $M=4$ and $M=8$ can be written as
\begin{eqnarray*}
\xi_4\;=\;\frac{1}{\sqrt{2}}\left(
\begin{array}{rr}
\xi_2\; & -\xi_2 \\ \xi_2 \;& \xi_2
\end{array}
\right)\;=\;
\frac{1}{2}\left(
\begin{array}{rrrr}
1  & -1 & -1 & 1\\
1  & 1 & -1 & -1 \\
1 & -1 & 1 & -1\\
1 & 1 & 1 & 1 
\end{array}
\right)
\end{eqnarray*}

\noindent and
\begin{eqnarray*}
\xi_8\;&=&\;\frac{1}{2}
\left(
\begin{array}{rrrr}
\xi_2  & -\xi_2 & -\xi_2 & \xi_2\\
\xi_2  & \xi_2 & -\xi_2 & -\xi_2 \\
\xi_2 & -\xi_2 & \xi_2 & -\xi_2\\
\xi_2 & \xi_2 & \xi_2 & \xi_2 
\end{array}\right)\hspace{20pt}{\rm etc.}
\end{eqnarray*}

\noindent where each matrix is written with respect to the ONB consisting of $\sqrt{M}{}^{(M)}\delta^m$,with $m=0,1,\ldots, M-1$. \\
Due to the property (\ref{Eq:IntertwiningUnitary}) one can immediately see that the family of $\{\xi_M\}_M$ can be used to construct an invertible isometric map between $\sqcup_M\mathcal{H}_M^{(i)}\big/\sim$  for $i=1,2$. By continuity, it extends to a unique unitary map
\begin{eqnarray}
\xi_\infty\;:\;\mathcal{H}_\infty^{(1)}\;\longrightarrow\;\mathcal{H}_\infty^{(2)}
\end{eqnarray}

\noindent which satisfies
\begin{eqnarray}
\xi_\infty I_M^{(1)}\;=\;I^{(2)}_M \xi_M
\end{eqnarray}
\noindent with the canonical embedding maps (\ref{Eq:CanonicalEmbeddingMaps}).

\subsection{Discretisation of operators \& RG flow}
\label{section:3_D-DiscOfOper}

At this point, we consider a discretisation of the fundamental operators (\ref{Eq:FundamentalWeylOperators}) for both cases $i=1,2$ and investigate their RG flow. For completeness (and to highlight why we are interested in the Weyl operators) we will also discuss at the end the RG fixed points of the non-exponentiated momentum operator.

For $i=1,2$, we define discrete versions of the basic operators via 
\begin{eqnarray}\label{Eq:Discretisation_01}
\begin{aligned}
(\hat{U}_M^{(i)}f)_m\;&:=\;\exp\left(\frac{2\pi i m}{M}\right) \,f_m\\[5pt]
(\hat{T}(l)_M^{(i)}f)_m\;&=\;\left\{
\begin{array}{ll}
f_{m+Ml}& \qquad m+Ml<M\\[5pt]
f_{m+M(l-1)}& \qquad m+Ml\geq M
\end{array}
\right.
\end{aligned}
\end{eqnarray}
\noindent where $l$ can take on the discrete values
\begin{eqnarray}
l\;\in\; \left\{0,\,\frac{1}{M},\,\frac{2}{M},\;\ldots,\;\frac{M-1}{M}\right\}.
\end{eqnarray} 

\noindent Note that the discretised operators satisfy
\begin{eqnarray}\label{Eq:Weyl_Relations_Discrete_01}
\hat{U}_M^{(i)}\,\hat{T}(l)_M^{(i)}\,(\hat{U}_M^{(i)})^\dag\;=\;e^{2\pi i l}\hat{T}(l)_M^{(i)},
\end{eqnarray}

\noindent which mimics the continuum algebra (\ref{Eq:Weyl_Relations_Continuum}), the difference being that $l$ can take on only discrete values. Indeed, each of the $\mathcal{H}_M^{(i)}$ thus becomes a representation space of a subalgebra of the continuum algebra.


\subsubsection{Case $i=1$}

It can be shown straightforwardly that both operators $\hat{U}_M^{(1)}$ and $\hat{T}(l)_M^{(1)}$ defined by (\ref{Eq:Discretisation_01}) are strongly cylindrically consistent for $i=1$, i.e.~they satisfy
\begin{eqnarray*}
\hat{U}_{2M}^{(1)}I^{(1)}_{M\to 2M}\;&=&\;I^{(1)}_{M\to 2M}\hat{U}_{M}^{(1)},\\[5pt]
\hat{T}(l)_{2M}^{(1)}I^{(1)}_{M\to 2M}\;&=&\;I^{(1)}_{M\to 2M}\hat{T}(l)_{M}^{(1)}.
\end{eqnarray*}

\noindent with respect to the embedding maps (\ref{Eq:EmbeddingMap01}). Therefore, they determine unitary continuum operators $\hat{U}_\infty^{(1)}$ and $\hat{T}(l)_\infty^{(1)}$ on the continuum Hilbert space $\mathcal{H}_\infty ^{(1)}$. Note that $l$ can take on all values
\begin{eqnarray}
l\;\in\;P\;=\;\left\{\frac{m}{M}\;\Big\vert\;M=2^L,\,m=0,\ldots,M-1\right\},
\end{eqnarray}

\noindent which is not all of $[0,1)$, but lies densely in it. These continuum operators also straightforwardly satisfy the algebra
\begin{eqnarray}
\hat{U}_\infty^{(i)}\,\hat{T}(l)_\infty^{(i)}\,(\hat{U}_\infty^{(i)})^\dag\;=\;e^{2\pi i l}\hat{T}(l)_\infty^{(i)}
\end{eqnarray}

\noindent The continuum limit therefore appears as a representation of a subalgebra of the original algebra (\ref{Eq:Weyl_Relations_Continuum}), in that only $l\in P\subset[0,1)$ are allowed.

\subsubsection{Case $i=2$}

The situation is more involved in the case of $i=2$. Notably, the operators $\hat{T}_M^{(2)}$ are also strongly cylindrically consistent with respect to the embedding maps $I^{(2)}_{M\to 2M}$ defined in (\ref{Eq:EmbeddingMap02}). However, the same cannot be said for the $\hat{U}_M^{(2)}$. Upon defining the bilinear forms
\begin{eqnarray}
U_M^{(2)}(m,n)\;:=\;\langle {}^{(2)}e_M^m,\,\hat{U}_M^{(2)}{}^{(2)}e_M^n\rangle
\:=\;e^{2\pi i\frac{m}{M}}\delta_{mn},
\end{eqnarray}

\noindent we find that
\begin{eqnarray*}
&& U^{(2)}_{2M}(I_{M\to2M} {}^{(2)}e^m_M, I_{M\to2M} {}^{(2)}e^n_M)\\[5pt]
&=&\frac{1}{2}\Big(U_{2M}^{(2)}(2m,2n)+U_{2M}^{(2)}(2m+1,2n)\\[5pt]
&&\quad +U_{2M}^{(2)}(2m,2n+1)+U_{2M}^{(2)}(2m+1,2n+1)\Big)\\[5pt]
&=&\;\frac{1}{2}\left(e^{2\pi i\frac{m}{M} }+e^{2\pi i\frac{m+1/2}{M} }\right)\delta_{mn}\\[5pt]
&\neq &\;U_M^{(2)}(m,n).
\end{eqnarray*}

\noindent Hence, the $\hat{U}_M^{(2)}$ are not even weakly cylindrically consistent in the sense of bilinear forms. Therefore, they do not define a continuum bilinear form on $\mathcal{H}_\infty^{(2)}$. \\
However, we can use the RG flow to construct a proper continuum operator:

{\bf Claim:} The RG flow, as defined in (\ref{RG_limit_operator}), of embedding map $I^{(2)}_{M\to2M}$ from (\ref{Eq:EmbeddingMap02})  starting from ${}^{(o)}U_M:=U^{(2)}_M$ flows into the weakly cylindrical consistent family:
\begin{align}
{}^{(\infty)}U_M(m,n)=e^{2\pi i\frac{m+1/2}{M}}\delta_{mn}\frac{\sin (\pi/M)}{\pi/M}
\end{align}

{\it Proof.} Explicitly, the RG flow equations read as follows:
\begin{eqnarray*}
{}^{(k+1)}U_M(m,n)\;&:=&\;\frac{1}{2}\sum_{\delta_0,\delta_1=0}^1
{}^{(k)}U_{2M}(2m+\delta_0,2n+\delta_1),\\[5pt]
{}^{(0)}U_M(m,n)\;&:=&\;U_M^{(2)}(m,n)\;=\;e^{2\pi i\frac{m}{M}}\delta_{mn}.
\end{eqnarray*}

\noindent From the form of the RG equations, we can see that if ${}^{(k)}U_M(m,n)\sim \delta_{mn}$, then also  ${}^{(k+1)}U_M(m,n)\sim \delta_{mn}$, so we find that
\begin{eqnarray*}
&&{}^{(k+1)}U_M(m,n)\\[5pt]
&&\quad\;=\;
\frac{1}{2}\left({}^{(k)}U_{2M}(2m,2n)+{}^{(k)}U_{2M}(2m+1,2n+1)\right)
\end{eqnarray*}

\noindent and hence, iterating this:
\begin{eqnarray*}
{}^{(k+p)}U_M(m,n)\:=\;2^{-p}\sum_{s=0}^{2^p-1}{}^{(k)}U_{2^pM}(2^pm+s,2^pn+s)\,\delta_{mn}.
\end{eqnarray*}

\noindent This allows us to explicitly compute the $p$-th step of the RG flow as
\begin{eqnarray}\label{Eq:RG_Weyl_02}
{}^{(p)}U_M(m,n)
\:=\:
2^{-p}\sum_{s=0}^{2^p-1}e^{2\pi i (m+2^{-p}s) /M}\;\delta_{mn}.
\end{eqnarray}

\noindent Taking the limit $p\to\infty$, the sum turns into an integral, and we get
\begin{eqnarray*}
{}^{(\infty)}U_M(m,n)\;&=&\;M\int_0^{1/M}dx\;e^{2\pi i(\frac{m}{M}+x)}\;\delta_{mn}\\[5pt]
&=&\:
e^{2\pi i\frac{m+1/2}{M}}\;\delta_{mn}\;\frac{\sin(\pi/M)}{\pi/M}.
\end{eqnarray*}

\noindent Hence, the RG flow has a fixed point, and the resulting limit partial operators define a continuum bilinear form $U_\infty^{(2)}$, given by
\begin{eqnarray}\label{Eq:BilinearFormUnitary_I=2}
U_\infty^{(2)}\left(I_M^{(2)}e_M^m, I_M^{(2)}e_M^n\right)\;:=\;{}^{(\infty)}U_M(m,n).
\end{eqnarray}

{\bf Claim:} This bilinear form on $\mathcal{H}^{(2)}_\infty$ can be extended to a unitary operator $\hat{U}_\infty^{(2)}$ such that
\begin{eqnarray}
\langle\,I_M^{(2)}e_M^m,\;\hat{U}_\infty^{(2)}I_M^{(2)}e_M^n\rangle\;=\;U_\infty^{(2)}\left(I_M^{(2)}e_M^m, I_M^{(2)}e_M^n\right).
\end{eqnarray}

{\it Proof.} This is shown in appendix \ref{Appendix:Extension_Biform_to_Operator}.\\ 

Here something peculiar happens: the operator $\hat{U}_\infty^{(2)}$ is indeed unitary, but its partial operators $\hat{U}_{\infty,M}:=(I_M^{(2)})^\dag\hat{U}_\infty^{(2)}I_M^{(2)}$ are not! Indeed, it is the additional factor of $\sin(\pi/M)/(\pi/M)$ which spoils the unitarity. This factor arose during the computation of the fixed point of the RG flow. It is a consequence of the fact that the RG fixed point operator $\hat{U}_\infty^{(2)}$ maps each vector $I_M^{(2)}e_M^m$ out of $\mathcal{D}_\infty^{(2)}$, i.e.~the image can not be written as an element of any individual $I_M^{(2)}(\mathcal{H}_M)$, but has to be expressed by an infinite sum. Its projection onto any $I_M^{(2)}(\mathcal{H}_{M'})$ will therefore reduce the norm of the vector, causing the appearance of the additional factor, which is of norm less than one. \\

For this example, we are in the advantageous position that $U_\infty^{(2)}, T_\infty^{(2)}$ already satisfy the correct algebraic relations (\ref{Eq:Weyl_Relations_Continuum}).\\
For the sake of completeness, we shall also test the prescription of subsection \ref{section:2-5_ModRG} which guarantees {\it a priori} that its fixed points will restore the correct Weyl-relation in the continuum. For this purpose, we start with ${}^{(o)}U_M$ and ${}^{(o)}T_M$ and study the flow of (\ref{RG_flow_consistent_algebra}). For ${}^{(o)}T_M$, one checks easily that it is already at its fixed point.\\
For ${}^{(o)}U_M$ the modified RG equations read explicitly:
\begin{align}
&{}^{(n+1)}U_M=e^{-2\pi\,i\,l}\times\\
&\lim_{N\to \infty}I^\dagger_{M\to 2^NM}{}^{(n)}T_{2^N M}(-l){}^{(n)}U_{2^N M}{}^{(n)}T_{2^N M}(l)I_{M\to 2^N M}\nonumber
\end{align}
After one step, we obtain:
\begin{align*}
&{}^{(1)}U_M(m,n)=\\
&=\lim_{N\to\infty}\frac{e^{-2\pi i l}}{2^N}\sum_{s,s'=0}^{2^N-1}{}^{(o)}U_{2^NM}(2^Nm+s+2^NMl,\\
&\hspace{150pt}2^Nn+s'+2^NMl)=\\
&=\lim_{N\to\infty} \frac{e^{-2\pi i l}}{2^N}\sum_{s,s'}\delta_{mn}\delta_{s,s'} e^{2\pi i (2^Nm+2^NMl+s)/(2^NM)}\\
&= \lim_{N\to\infty} 2^{-N} \sum_{s=0}^{2^N-1}e^{2\pi i (m+2^{-N} s)/M} \delta_{mn}\\
&=e^{2\pi i \frac{m+1/2 }{M}}\delta_{mn}{\rm sinc}(\pi/M)
\end{align*}
as was already shown for (\ref{Eq:RG_Weyl_02}).\\
Indeed, this is already a fixed point of (\ref{RG_flow_consistent_algebra}), since ${}^{(1)}U_\infty(m,n)={}^{(o)}U_\infty(m,n)$ for all $m,n$, i.e. the bilinear forms in the continuum agree and only those enter the flow of (\ref{RG_flow_consistent_algebra}).

\subsubsection{Momentum operators}\label{Sec:MomentumOperators}

Additionally to the translation operators $\hat{T}(l)$, on the continuum Hilbert space $ L^2([0,1],dx)$ the infinitesimal version, i.e.~the momentum operator 
\begin{eqnarray}
\hat{p}f\:=\:-i\frac{d}{dl}_{\vert l=0}\hat{T}(l)f\;=\;-i\frac{df}{dx}
\end{eqnarray}

\noindent exists. This operator is self-adjoint an unbounded, i.e.~it is only defined on a dense domain of $\mathcal{H}$, which, however, includes $\mathcal{D}$ \cite{reed1975ii}. The momentum operator is essentially a differentiation w.r.t.~$x$, and neither discretisation $i=1,2$ straightforwardly allows such a differentiation, since $x$ can take only discrete values. Still, one can construct a finite difference operator and investigate its properties.

On the Hilbert spaces $\mathcal{H}_M^{(i)}$, $i=1,2$, the differentiation can be approximated by finite differences. For this, we assume periodic boundary conditions, and denote, for any $f\in \mathcal{H}_M^{(i)}$,
\begin{eqnarray}\label{Eq:PeriodicBoundaryCondition}
f_M\;:=\;f_0.
\end{eqnarray} 

\noindent With this we define sesquilinear forms $\partial_M{(i)}$, $i=1,2$, by
\begin{eqnarray}
(\partial_M)_{m,n}^{(i)}\;:=\;2M(\delta_{m,n+1}-\delta_{m,n-1}).
\end{eqnarray}
\noindent Checking (\ref{weak_cyl_consistency}) cylindrical consistency first for $i=1$, we get that 
\begin{eqnarray*}
\partial_{2M}^{(1)}\big(I_{M\to 2M}^{(1)}e_m^M,\,I_{M\to 2M}^{(1)}e_n^M\big)
\;&=&\;
\partial_{2M}^{(1)}\big(e_{2m}^{2M},\,e_{2n}^{2M}\big)\\[5pt]
&=&\:
4M\big(\delta_{2m,2n+1}-\delta_{2m,2n-1}\big)\\[5pt]
&=&\;
0.
\end{eqnarray*}
\noindent So this series of sesquilinear forms ist not cylindrically consistent. Moreover, by the above calculation we have confirmed that, in the RG flow, the family of forms becomes the zero form, after only one RG step.\\

Turning to $i=2$ next, one can show that
\begin{eqnarray*}
&&
\partial_{2M}^{(2)}\big(I_{M\to 2M}^{(2)}e_m^M,\,I_{M\to 2M}^{(2)}e_n^M\big)\\[5pt]
&&\qquad\;=\;
\partial_{2M}^{(2)}\left(\frac{e_{2m}^{2M}+e_{2m+1}^{2M}}{\sqrt{2}},\,\frac{e_{2n}^{2M}+e_{2n+1}^{2M}}{\sqrt{2}}\right)\\[5pt]
&&\qquad\;=\,
2M\big(\delta_{2m+1,2n}-\delta_{2m-1,2n}+\delta_{2m+2,2n}-\delta_{2m,2n}\\[5pt]
&&\qquad\;\;
+\delta_{2m+1,2n+1}-\delta_{2m-1,2n+1}+\delta_{2m+2,2n+1}-\delta_{2m,2n+1}
\big)\\[5pt]
&&\qquad\; =\partial_M^{(2)}(e_m^M,e_n^M)
\end{eqnarray*}
\noindent Hence, this family of sesquilinear forms is already cylindrically consistent, and therefore straightforwardly defines a sesquilinear form
\begin{eqnarray}\label{Eq:MomentumForm_2}
\partial_\infty^{(2)}\;:\;\mathcal{D}_\infty^{(2)}\times \mathcal{D}_\infty^{(2)}\;\to\;\mathbb{C}.
\end{eqnarray}
\noindent For $f,g\in\mathcal{H}_M^{(2)}$, one has that
\begin{eqnarray*}
\langle f,\,\partial_Mg\rangle
\;&=&\;\sum_{m=0}^{M-1}\overline{f_m}(g_{m+1}-g_{m-1})
\;=\;-\;{\langle \partial_M f,\,g\rangle},
\end{eqnarray*}
\noindent and hence $\partial_M^{(2)}$ is anti-Hermitean on $\mathcal{H}_M^{(2)}$, due to our periodic boundary conditions (\ref{Eq:PeriodicBoundaryCondition}). Therefore, also $\partial_\infty^{(2)}$ is.
The situation of the sesquilinear form (\ref{Eq:MomentumForm_2}) is interesting, since, as one can straightforwardly show, it does \emph{not} satisfy (\ref{Eq:ConditionBilinearFormIsAnOperator}), and hence does not define a densely defined operator on $\mathcal{D}_\infty^{(2)}\subset \mathcal{H}_\infty^{(2)}$. However, we know that $\mathcal{H}_\infty^{(2)}\simeq L^2(S^1)$, and $\partial_\infty^{(2)}$ has a connection with the sesquilinear form
\begin{equation}\label{Eq:StandardMomentumBilinearForm}
\partial(\psi,\phi)\;:=\;i\langle\psi,\hat{p}\phi\rangle
\end{equation}
\noindent for $\psi,\phi\in\,\mathcal{D}=C^\infty(S^1)$. Indeed, $\partial_\infty^{(2)}$ can be extended continuously (in the nuclear topology on $\mathcal{D}$) to a sesquilinear form on
\begin{eqnarray}\label{Eq:DomainsOfDefinition}
\overline{\mathcal{D}}\;:=\mathcal{D}+\mathcal{D}_\infty^{(2)},
\end{eqnarray}
\noindent the restriction of which on $\mathcal{D}$ then agrees with  (\ref{Eq:StandardMomentumBilinearForm}). In that sense, the $\partial_M$ are the correct discretisation of the continuum derivation operator. 
This shows that, even though the collection of partial Hilbert spaces correctly approximate the continuum, and certain physical observables can be defined properly on the continuum, the issue of finding the correct dense domain of definition for the resulting operator is not necessarily addressed or solved by the RG flow procedure. In the specific example of the momentum operator, this has to do with the fact that the discretisation and some aspects of the continuum operators (such as differentiation) are at odds with one another. This can happen in a strong sense, as with $i=1$, where the naive discretisation of the continuum operators immediately flows to the zero operator, or in the more subtle case $i=2$, where the naive discretisation leads to a continuum form, for which the correct dense domain of definition still has to be found, in order to make it into an operator on the continuum Hilbert space.

\subsection{Comparison of continuum limits}\label{Sec:COmparisonOfContinuumLimitsOfQMonnaCircle}

So far, we have considered two different discretisations of quantum mechanics on the circle, and considered their respective continuum limits. In both cases we obtained continuum Hilbert spaces $\mathcal{H}_\infty^{(i)}$, and (in one case only through renormalization) a consistent family of discretised operators $\{\hat{U}^{(i)}_M,\,\hat{T}(l)^{(i)}_M\}_M$, $i=1,2$, which extended to operators $\hat{U}_\infty^{(i)},\,\hat{T}(l)_\infty^{(i)}$ on the respective continuum limit Hilbert spaces.

These two discretisations are not equivalent, in the following sense: There is a family of isometries
\begin{eqnarray}
\xi_M\;:\;\mathcal{H}_M^{(1)}\;\longrightarrow\;\mathcal{H}_M^{(2)}
\end{eqnarray}

\noindent which intertwine the embedding maps, i.e.~satisfy
\begin{eqnarray}
\xi_{2M}\circ I^{(1)}_{M\to 2M}\;=\;I^{(2)}_{M\to 2M}\circ \xi_M,
\end{eqnarray}

\noindent and therefore extend to an isometry $\xi_\infty$ between the two continuum Hilbert spaces. However, as a quick calculation shows, these isometries do not intertwine the discrete operators $\hat{U}_M^{(i)}$, $\hat{T}_M^{(i)}$.\footnote{In the case of the $\hat{U}_M^{(i)}$ this can easily be seen by the fact that one of the two is unitary, the other is not. It is, however, also true for the $\hat{T}(l)^{(i)}_M$. } As a result, one has e.g.~
\begin{eqnarray}
\xi_\infty \hat{U}_\infty^{(1)}\xi_\infty^{-1}\;\neq\;\hat{U}_\infty^{(2)}.
\end{eqnarray}

\noindent Indeed, there is \emph{no} unitary equivalence between the two continuum Hilbert spaces and the respective operators. This can be seen easily e.g.~by observing that every $e_m^M$ is a normalisable Eigenvector of $\hat{U}^{(1)}_\infty$, while $\hat{U}_\infty^{(2)}$ has $U(1)$ as its continuous spectrum. Also, since $\mathcal{H}^{(2)}_\infty\simeq L^2(S^1)$ with the standard quantisation of $\hat{U}$ and $\hat{T}(l)$ (see section \ref{section:3_1_Continuum}), the family ${}^{(2)}\hat{T}_\infty(l)$ can be extended to $e^{2\pi il}\in U(1)$, which is a strongly continuous 1-parameter family. On the other hand, while $l_n:=2^{-n}$ converges to $0$, the vector ${}^{(1)}\hat{T}_\infty(l_n)\big(e_1^0\big)$ is orthogonal to $e_1^0$, therefore the family ${}^{(2)}\hat{T}_\infty(l)$ can not be extended continuously to real $l$. In particular, there is no self-adjoint $\hat{p}$ such that
\begin{eqnarray}
{}^{(1)}\hat{T}_\infty(l)\;=\;e^{il\hat{p}}.
\end{eqnarray}

\noindent It should be noted that this makes this type of quantisation conceptually very similar to Loop Quantum Gravity, and (if one replaces the circle by the real line), goes by the name \emph{Polymer Quantum Mechanics} (see \cite{Corichi:2007tf}). \\

The fact that both quantisations feature operators with different spectra, highlights the fact that in general several unitary inequivalent representations of the same algebra exist. And the method of inductive limit can not necessarily serve as a procedure to physically distinguish them. However, it can help in classifying them, as they can all be obtained as the fixed points of the RG flow for one and the same choice of embedding map and different suitable initial starting points.

\subsection{Weak equivalence of $i=1,2$}\label{Sec:WeakEquivalenceOfi=1and2}

Here we show that although the two discretisations for $i=1,2$ are unitarily inequivalent, they are at least weakly equivalent in the sense that they approximate the same continuum physics. 

To show this, we construct the embedding maps $\phi_M^{(1)}:\mathcal{H}_M^{(i)}\to \mathcal{D}'$, i.e.~of discretised states into distributions on the unit circle:
\begin{eqnarray}\label{Eq:DistributionalEmbedding_01}
\phi_M^{(1)}e_M^m
\;&:=&\;
\delta_{m/M} \\[5pt]\label{Eq:DistributionalEmbedding_02}
\phi_M^{(2)}e_M^m
\;&:=&\;
\sqrt{M}\chi_{[\frac{m}{M},\frac{m+1}{M})},
\end{eqnarray}

\noindent where $\delta_x$ denotes the delta-distribution at $x\in[0,1)$, and $\chi_{[a,b)}$ the characteristic function on the interval $[a,b)$. 
It turns out that we cannot compare both $\mathcal{H}^{(i)}_\infty$ directly, as only one embedding allows to be extendible to the corresponding Hilbert space:

{\bf Claim:} For both discretisations, the maps $\phi^{(i)}_M$ are faithful embeddings maps. Also, the case $i=2$ is regular and separating, while $i=1$ is not regular.

{\it Proof.} (i) We need to verify (\ref{Eq:FaithfulEmbedding_02}) and since $I_{M\to 2M}^{(1)}e_M^m=e_{2M}^{2m}$, it is clear that (\ref{Eq:DistributionalEmbedding_01}) commutes with the embedding maps for $i=1$. For $i=2$ we have:

\begin{eqnarray*}
\phi_{2M}^{(2)}I_{M\to 2M}^{(2)}e_M^m
\;&=&\;
\phi_{2M}^{(2)}\left(\frac{e_{2M}^{2m}+e_{2M}^{2m+1}}{\sqrt{2}}\right)\\[5pt]
&=&\:\frac{\sqrt{2M}}{\sqrt{2}}\left(\chi_{[\frac{2m}{2M},\frac{2m+1}{2M})}+\chi_{[\frac{2m+1}{2M},\frac{2m+2}{2M})}\right)\\[5pt]
&=&\;\sqrt{M}\chi_{[\frac{m}{M},\frac{m+1}{M})}
\;=\;
\phi_M^{(2)}e_M^m.
\end{eqnarray*}

\noindent Hence, both $\phi_M^{(i)}$ define faithful embeddings, since they evidently are injective.\\

To show the regularity for $i=2$, we again take $\psi=\sum_{m}a_m e_M^m$ and a function $f\in \mathcal{D}$, and remark that we have
\begin{eqnarray}
\langle \phi_M^{(2)}(\psi), \,f\rangle
\;&=&\:
\sqrt{M}\sum_ma_m\int_{m/M}^{(m+1)/M}dx f(x),
\end{eqnarray}
\noindent and hence, by the Cauchy-Schwartz inequality:
\begin{eqnarray*}
\vert\langle \phi_M^{(2)}(\psi), \,f\rangle\vert^2
\;&\leq &\:
M\left(\sum_m\vert a_m\vert^m\right)\sum_m \int_{m/M}^{(m+1)/M}dx \vert f(x)\vert^2
\\[5pt]
&\leq &\;\Vert\psi\Vert^2\,M\sum_m \int_{m/M}^{(m+1)/M}dx f_\text{max}^2\\[5pt]
&=&\:\Vert\psi\Vert^2f_\text{max}^2.
\end{eqnarray*}

\noindent To see that $\phi_\infty^{(1)}$ can not be regular, it is enough to consider the Cauchy sequence
\begin{eqnarray}
\psi_n\;:=\;\sum_{1\leq k\leq n}\frac{1}{k}e_k,
\end{eqnarray}

\noindent where $\{e_k\}$ denotes an ONB lying in $\mathcal{D}_\infty$ such that $e_k=e_{M}^m$ for some $M,m$, i.e.~ 
\begin{equation}
\phi_\infty^{(1)}(e_k)\;=\;\delta_{x_k}
\end{equation}

\noindent with $x_k$ being a sequence in $P$ without any element being hit more than once. It is clear that $\phi_\infty^{(1)}(\psi_n)$ does not converge in $D'$, e.g.~by considering its action on the constant function:
\begin{equation}
\langle\phi_\infty^{(1)}(\psi_n),\,1\rangle\;=\;\sum_{1\leq k\leq n}\frac{1}{k},
\end{equation}

\noindent which diverges, while $\psi_n$ converges in $\mathcal{H}_\infty^{(1)}$. 

\noindent From this it follows that for both discretisation $i=1,2$ there exist faithful embeddings $\phi_\infty^{(i)}$ from $\mathcal{D}^{(i)}_\infty$ into the distributions $\mathcal{D}'$. It is also straightforward to show that the extension of $\phi_\infty^{(2)}$ to $\mathcal{H}_\infty^{(2)}$ is separating, i.e.~no element in $\mathcal{H}_\infty^{(2)}$ gets mapped to the zero distribution.\\

Thus, we can try to establish at least weak equivalency. We will do this now, by first confirm the conditions of definition \ref{Def:approx_continuum}.\\

{\bf Claim:} Both faithful embeddings $\phi^{(i)}_\infty$ embed densely, i.e.~ $\phi_\infty^{(i)}(\mathcal{D}_\infty^{(i)})\subset \mathcal{D}'$ is dense in the weak-$*$-topology on $\mathcal{D}'$.

{\it Proof.} For $i=2$ this is easy to show, since the image $\phi_\infty^{(2)}(\mathcal{D}_\infty^{(2)})$ is actually the space of (regular\footnote{They constitute not all characteristic functions on $[0,1)$, since the intervals are only allowed to have endpoints which are of the form $m/2^L$. But since these sets are a basis for the topology on $[0,1)$, the statement still holds.}) piecewise constant functions on $[0,1)$, which are well-known to be dense in $C^0([0,1))$ in the  uniform topology, which is stronger than the weak-$*$-topology, and $C^0([0,1))$ is dense in $\mathcal{D}'$ in the weak-$*$-topology, so we are done. Indeed, we have shown that one can canonically identify
\begin{eqnarray}
\phi_\infty^{(2)}(\mathcal{H}_\infty^{(2)})\;=\;L^2([0,1))\;=\;\mathcal{H}\;\subset\;\mathcal{D}'.
\end{eqnarray}

\noindent For $i=1$, it is enough to show that it is dense\footnote{The reader is reminded that ``$X$ being dense in $Y$'' does \emph{not} imply that $X\subset Y$, it only means that $Y\subset \overline{X}$.} in $\mathcal{D}$. Hence, let $f\in\mathcal{D}$. It is now enough to show that there is a sequence $\psi_n$ in $\phi_\infty^{(1)}(\mathcal{D}_\infty^{(1)})$ which converges to $f$ in the weak-$*$-topology, i.e.~that, for every $g\in\mathcal{D}$ we have
\begin{eqnarray}
\lim_{n\to\infty}\langle \psi_n,\,g\rangle\;=\;\langle f,\,g\rangle\;=\;\int_0^1dx\,f(x)g(x).
\end{eqnarray}

\noindent For this we choose
\begin{eqnarray}
\psi_n\;=\;\frac{1}{2^n}\sum_{m=0}^{2^n-1}f\left(\frac{m}{2^n}\right)\,\delta_{\frac{m}{2^n}}.
\end{eqnarray}

\noindent As one can see from (\ref{Eq:DistributionalEmbedding_01}), each $\psi_n$ is in $\phi_M^{(1)}(\mathcal{H}_M^{(1)})$ for $M=2^n$. We then have
\begin{eqnarray}
\langle \psi_n,\,g\rangle
&=&\;
\frac{1}{2^n}\sum_{m=0}^{2^n-1}f\left(\frac{m}{2^n}\right)\,g\left(\frac{m}{2^n}\right).
\end{eqnarray}

Since Riemann sums approximate $L^2$-integrals, this clearly converges to $\langle f,\,g\rangle$ as $n\to\infty$, and we are done. Note that the sequence of the $\psi_n$ also converges in the stronger $L^2$-topology, albeit always to the zero vector in $\phi_\infty^{(1)}(\mathcal{H}_\infty^{(1)})$.\\

It remains to show that not only the Hilbert spaces are weakly equivalent, but also the operators $U^{(i)}_\infty$ and $T_\infty^{(i)}(l)$.

{\bf Claim:} The operators $U^{(i)}_\infty$ and $T_\infty^{(i)}(l)$ obey (\ref{Eq:OperatorsApproximateTheContinuum}), i.e.~the respective continuum limits are weakly equivalent. \\

\emph{Proof:} In order to show their weak equivalence, one needs to show that they are both weakly equivalent to the original continuum physics, i.e.~one has
\begin{eqnarray}
\langle \phi_\infty^{(i)}(\hat{U}_\infty^{(i)}\psi),\;f\rangle\;&=&\;\langle \phi_\infty^{(i)}(\psi),\;\hat{U}^\dag f\rangle\\[5pt]
\langle \phi_\infty^{(i)}(\hat{T}(l)_\infty^{(i)}\psi),\;f\rangle\;&=&\;\langle \phi_\infty^{(i)}(\psi),\;\hat{T}(l)^\dag f\rangle
\end{eqnarray}

\noindent for either $i=1,2$, all $\psi\in\mathcal{D}_A^{(i)}$, all $f\in \mathcal{D}$, where $\mathcal{D}_A^{(i)}$ is an invariant domain of definition of both $\hat{U}_\infty^{(i)}$ and $\hat{T}_\infty^{(i)}(l)$, which $\phi_\infty^{(i)}$ can be extended to, and where $\hat{U}$ and $\hat{T}(l)$ are the usual continuum operators (\ref{Eq:FundamentalWeylOperators}).\\

First we show the case $i=1$. Here we choose $\mathcal{D}_A^{(1)}=\mathcal{D}_\infty^{(1)}$, which is left invariant by both operators $\hat{U}_\infty^{(1)}$ and $\hat{T}_\infty^{(1)}(l)$. Due to linearity, it is enough to show the claim for $\psi=e_M^m$, i.e.~with (\ref{Eq:DistributionalEmbedding_01}):
\begin{eqnarray}
\phi_\infty^{(1)}(\psi)\;=\;\delta_x
\end{eqnarray}
\noindent for some $x\in P$. Thus by anti-linearity, we get
\begin{eqnarray}
\phi_\infty^{(1)}(\hat{U}_\infty^{(1)}\psi)\;=\;\phi_\infty^{(1)}(e^{2\pi i x}\psi)\;=\;e^{-2\pi i x}\delta_x
\end{eqnarray}

Thus we have
\begin{eqnarray*}
\langle \phi_\infty^{(1)}(\hat{U}_\infty^{(1)}\psi),\;f\rangle
\;&=&\;
e^{-2\pi i x}f(x)\;
=\;\langle \delta_x,\;\hat{U}^\dag f\rangle,
\end{eqnarray*}
\noindent which shows the claim for $\hat{U}^{(1)}_\infty$. For $\hat{T}_\infty^{(l)}$ the claim follows directly from the translation invariance of the integral. \\

For $i=2$, we note that the operators $\hat{U}_\infty^{(2)}$ and $\hat{T}_\infty^{(2)}(l)$ are bounded and can be extended to all of $\mathcal{H}_\infty^{(2)}$, which is also true for $\phi_\infty^{(2)}$, since the faithful embedding is regular and separating. Therefore, we are free to choose any dense invariant domain. Since $\phi_\infty^{(2)}(\mathcal{H}_\infty^{(2)})=\mathcal{H}$ (seen as subspace of $\mathcal{D}'$), we can choose $\mathcal{D}_\infty^{(2)}=(\phi_\infty^{(2)})^{-1}(\mathcal{D})$. However, it can be shown that the action of $\hat{U}_\infty^{(2)}$ on $(\phi_\infty^{(2)})^{-1}(\mathcal{D})$ coincides with the pull-back of the continuum operator $\hat{U}$. The proof is given in appendix \ref{Appendix:Extension_Biform_to_Operator}. The same can be trivially shown for the translation operator. Thus, the claim is shown.

\section{Conclusion}
\label{section:4_Conclusion}
In this article  we have studied different properties of the Hamiltonian Renormalisation.
We showed that there is (at least conceptually) a redundancy when looking at embedding maps and initial discretisations. One of both may be fixed without loosing information. This could in principle motivate to pick once and for all an embedding map for which one has control over the numerical approximations and look at the fixed points obtained from several initial discretisations.\\
Of course, different initial discretisations will in general flow into different fixed points and we did not establish criteria to identify whether those turn out to be trivial or physically interesting - this remains to be checked a posteriori just as in the path-integral renormalisation framework. However, in contrast to there, we could formulate a modified RG flow, which (in suitable situations) will drive initially discretised algebras of operators\footnote{It is worth to mention that in most physically interesting situations the algebraic relations {\it must} be violated at the discrete level.} to those fixed points which restore the correct continuum commutation relations (given that the fixed point theory is not trivial or divergent). This improves over the situation in covariant RG.\\
We checked these claims explicitly for the case of quantum mechanics on the circle: we proposed an initial discretisation and two embedding maps such that two relevant fixed points were found. In both cases exponentiated momentum and position operator fulfil the Weyl algebra in the continuum. Importantly, both fixed point theories are such that the operators are unitarily inequivalent, e.g. they have different spectra. This highlights that indeed there exist fixed points with different physical properties and further input is required to single a preferred one out.\\

Finally, we introduced the notion of ``weak equivalence'' between inductive limit-theories. Taking advantage of usually being interested in the distributions $\mathcal{D}'$ over the Hilbert space $\mathcal{H}\subset\mathcal{D}'$ as well, we established a condition under which even unitary inequivalent theories can be embedded into the same $\mathcal{D}'$ such that they both approximate the theory on $\mathcal{H}$.\\
This was again verified for the afore-mentioned test case. Thus, although the fixed point theories were different, in the sense of weak equivalence both are valid descriptions of the same system. This hints at even more hidden redundancy in the choice of initial data for investigations of the RG flow.\\

We hope that further research in this direction helps in identifying physically interesting theories for cases where the continuum QFT effects are yet unknown such as quantum gravity.\\

{\bf Acknowledgement:} The authors acknowledges support by the German Research Foundation (DFG) under Germany\'s Excellence Strategy - EXC 2121 ``Quantum Universe'' - 390833306. This work was partially funded by DFG-project BA 4966/1-2.

\appendix 

\section{An adapted ONB}\label{App:AdaptedONB}

Given a discrete system $\{\mathcal{H}_M,\,I_{M\to 2M}\}_M$ with finite-dimensional $\mathcal{H}_M$, there is a very useful orthonormal basis for the continuum Hilbert space $\mathcal{H}_\infty$, which we will use repeatedly in proofs in this article. 

For any $M$ we define an ONB adapted to $\mathcal{H}_M$ the following way: The first $\dim\mathcal{H}_M$ basis vectors are given by
\begin{eqnarray*}
e^n_\infty\;:=\;I_{M}e_M^n,\qquad n=1,\ldots,\dim\mathcal{H}_M.
\end{eqnarray*}

\noindent The next $\dim\mathcal{H}_{2M}-\dim\mathcal{H}_M$ vectors are constructed as follows: Complete the $I_{M\to 2M}e_M^n$ by $e^n_{2M}$ to an ONB of $\mathcal{H}_{2M}$ and define
\begin{eqnarray*}
e^n_\infty\;:=\;I_{2M}e_{2M}^n,\qquad n=\dim\mathcal{H}_M+1,\ldots,\dim\mathcal{H}_{2M}.
\end{eqnarray*} 

\noindent We repeat this successively for each $M'>M$, and obtain an ONB $\{e^n_\infty\}_n$ for $\mathcal{H}_\infty$ which has the property that the first $\dim\mathcal{H}_{M'}$ of these form an ONB of $I_{M'}(\mathcal{H}_{M'})$.

\section{A unitary operator on $\mathcal{H}_\infty^{(2)}$}
\label{Appendix:Extension_Biform_to_Operator}

We consider the bilinear form (\ref{Eq:BilinearFormUnitary_I=2}) given by
\begin{eqnarray*}
U_\infty^{(2)}\left(I_M^{(2)}e_M^m, I_M^{(2)}e_M^n\right)\;:=\;e^{
2\pi i\frac{m+1/2}{M}}\delta_{mn}\frac{\sin(\pi/M)}{\pi/M}.
\end{eqnarray*}

\noindent In order to show that this bilinear form can be extended to an operator, we use the unitary map $\phi_\infty^{(2)}:\mathcal{H}_\infty^{(2)}\to L^2([0,1))$ which is
given by
\begin{equation}
\phi_\infty^{(2)}I_Me_M^m\;=\;\sqrt{M}\,\chi_{[\frac{m}{M},\frac{m+1}{M})},
\end{equation}

\noindent i.e.~maps basis vectors to characteristic functions. We then find that the phase operator $\hat{U}$ (\Ref{Eq:FundamentalWeylOperators}) satisfies:
\begin{eqnarray*}
&&\langle \phi_\infty^{(2)}I_Me_M^m,\;\hat{U}\,\phi_\infty^{(2)}I_Me_M^n\rangle\\[5pt]
&&\quad \;=\;M\int_{0}^{1}dx \,e^{2\pi i x}\chi_{[\frac{m}{M},\frac{m+1}{M})}(x)\chi_{[\frac{n}{M},\frac{n+1}{M})}(x)\\[5pt]
&&\quad \;=\;\delta_{mn}\int_{m/M}^{(m+1)/M}dx\,e^{2\pi i x} \;=\;\frac{M \delta_{mn}}{2\pi i}\left(e^{2\pi i \frac{m}{M}}-e^{2\pi i \frac{m+1}{M}}\right)\\[5pt]
&&\quad\;=\;U_\infty^{(2)}\left(I_M^{(2)}e_M^m, I_M^{(2)}e_M^n\right)
\end{eqnarray*}

\noindent Hence, we see that the pull-back of the bilinear form defined by $\hat{U}$ to $\mathcal{H}_\infty^{(2)}$ via $\phi_\infty^{(2)}$ coincides with $U_\infty^{(2)}$. It therefore can be extended to an operator 
\begin{eqnarray}
\hat{U}_\infty^{(2)}\;:=\;\phi_\infty^{(2)}\,\hat{U}\,(\phi_\infty^{(2)})^{-1}.
\end{eqnarray}

\section{Proof of the theorem}
\label{proof_thm}

We prove theorem \ref{Theorem1}. So assume we are given (weakly) cylindrically consistent operators $\hat{A}_M$, $\hat{B}_M$, whose continuum bilinear forms $A_\infty$, $B_\infty$ define closable operators $\hat{A}_\infty$, $\hat{B}_\infty$ which have $\mathcal{D}_\infty$ as common invariant subspace. 

We use the adapted base $e_M^n$ from appendix \Ref{App:AdaptedONB}, and define:
\begin{eqnarray*}
{}^{(0)}C_M(m,n)\;&:=&\;\langle e_M^m,\,\hat{A}_M\hat{B}_Me_M^n\rangle\qquad m,n=1,\ldots \dim\mathcal{H}_M
\end{eqnarray*}

\noindent for all $M$ as the initial point of the RG flow defined by (\ref{RG_limit_operator}). Since the adapted basis satisfies
\begin{eqnarray*}
I_{M\to M'}e_M^n
\;=\;
&
e_{M'}^n,
\end{eqnarray*}
\noindent after $k$ steps the the flow results in 
\begin{eqnarray*}
{}^{(k)}C_M(m,n)
\;&=&\;
\langle e_{2^kM}^m,\,\hat{A}_{2^kM}\hat{B}_{2^kM}e_{2^kM}^n\rangle\\[5pt]
&=&\;
\sum_{l=1}^{L(k)}\langle e_{2^kM}^m,\,\hat{A}_{2^kM}e_{2^kM}^l
\rangle
\langle
e_{2^kM}^l,\,
\hat{B}_{2^kM}e_{2^kM}^n\rangle
\end{eqnarray*}

\noindent with $L(k)=\dim\mathcal{H}_{2^kM}$.

On $\mathcal{D}_\infty$ the product $\hat{C}_\infty:=\hat{A}_\infty\hat{B}_\infty$ exists, with the bilinear form 
\begin{eqnarray*}
C_\infty(m,n)\;&:=&\;\langle e_\infty^m,\,\hat{A}_\infty\hat{B}_\infty e_\infty^n\rangle\qquad m,n\in\mathbb{N}.
\end{eqnarray*}

\noindent Since by requirement both $\hat{A}_\infty$ and $\hat{B}_\infty$ are closable, we can form adjoints on $\mathcal{D}_\infty$, and get
\begin{eqnarray*}
C_\infty(m,n)
\;&=&\;
\langle (\hat{A}_\infty)^\dag e_\infty^m,\,\hat{B}_\infty e_\infty^n\rangle\\[5pt]
&=&\;\sum_{l=1}^\infty
\langle (\hat{A}_\infty)^\dag e_\infty^m,\,e_\infty^l\rangle\langle e_\infty^l,\,\hat{B}_\infty e_\infty^n\rangle\\[5pt]
&=&\;
\sum_{l=1}^\infty
\langle e_\infty^m,\,\hat{A}_\infty
e_\infty^l\rangle\langle e_\infty^l,\,\hat{B}_\infty e_\infty^n\rangle\\[5pt]
&=&\;
\lim_{k\to\infty}
\sum_{l=1}^{L(k)}
\langle e_\infty^m,\,\hat{A}_\infty
e_\infty^l\rangle\langle e_\infty^l,\,\hat{B}_\infty e_\infty^n\rangle,
\end{eqnarray*}

\noindent since $L(k)=\dim\mathcal{H}_{2^kM}$ tends to infinity. Now let $m,n\leq \dim\mathcal{H}_M=L(0)$, then by construction of our ONB we have
\begin{eqnarray*}
C_M(m,n)\;&=&\;C_\infty(m,n)\\[5pt]
&=&\;
\lim_{k\to\infty}
\sum_{l=1}^{L(k)}
\langle e_{2^kM}^m,\,\hat{A}_{2^kM}
e_{2^kM}^l\rangle\langle e_{2^kM}^l,\,\hat{B}_{2^kM} e_{2^kM}^n\rangle\\[5pt]
&=&\:
\lim_{k\to\infty} {}^{(k)}C_M(m,n).
\end{eqnarray*}

\noindent Hence, we have shown that the partial bilinear forms of the product of the continuum operators indeed coincide with the limit of the RG flow of the product of the partial operators. This finishes the proof.$\hfill \square$\\

\noindent It should be noted that it is easy to see that this proof straightforwardly extends to a product of $N$ operators. The crucial point is the appearance of multiple limits by introducing several resolutions of unity, which however clearly commute. \\

 We point out the necessity of both operators leaving the inductive-limit domain $\mathcal{D}_\infty$ invariant. This restriction is indeed unavoidable and cannot easily be lifted as one convinces himself with the following example: consider the momentum operator $\hat{p}$ from paragraph 3 of subsection \ref{section:3_D-DiscOfOper} whose domain of definition $\mathcal{D}$ does not agree with the inductive-limit domain $\mathcal{D}^{(2)}_\infty$ on which a weakly cylindrical consistent bilinear form was obtained (see \ref{Eq:DomainsOfDefinition}). Now, for $\hat A=\hat B:=\hat{p}$ albeit their product $\hat{A}\hat{B}=\hat p ^2$ exists on $\mathcal{D}$ it does not exist on $\mathcal{D}^{(2)}_\infty$ and fittingly also the RG flow of $C_M:=(\partial_M)^2$ diverges, invalidating the generality of theorem \ref{Theorem1} for cases where $\mathcal{D}\neq \mathcal{D}_\infty$.\\

\bibliography{bibliography}

\begin{thebibliography}{10}

\bibitem{GL:54}
M.~Gell-Mann and F.~Low, ``{Quantum electrodynamics at small distances},'' {\em
  Phys. Rev.}, vol.~95, pp.~1300--1312, 1954.

\bibitem{WK:74}
K.~Wilson and J.~B. Kogut, ``{The Renormalization group and the epsilon
  expansion},'' {\em Phys. Rept.}, vol.~12, pp.~75--199, 1974.

\bibitem{BB:01}
C.~Bagnuls and C.~Bervillier, ``Exact renormalization group equations: an
  introductory review,'' {\em Physics Reports}, vol.~348, p.~91–157, Jul
  2001.

\bibitem{GRS:14}
R.~Gurau, V.~Rivasseau, and A.~Sfondrini, ``Renormalization: an advanced
  overview,'' 2014.

\bibitem{Kad:66}
L.~Kadanoff, ``{Scaling laws for Ising models near T(c)},'' {\em Physics
  Physique Fizika}, vol.~2, pp.~263--272, 1966.

\bibitem{Kogut:79}
J.~B. Kogut, ``An introduction to lattice gauge theory and spin systems,'' {\em
  Rev. Mod. Phys.}, vol.~51, pp.~659--713, Oct 1979.

\bibitem{Cre:84}
M.~Creutz, {\em {Quarks, gluons and lattices}}.
\newblock Cambridge Monographs on Mathematical Physics, Cambridge, UK:
  Cambridge Univ. Press, 6 1985.

\bibitem{Nagy:14}
S.~Nagy, ``Lectures on renormalization and asymptotic safety,'' {\em Annals of
  Physics}, vol.~350, p.~310–346, Nov 2014.

\bibitem{Loll:19}
R.~Loll, ``Quantum gravity from causal dynamical triangulations: a review,''
  {\em Classical and Quantum Gravity}, vol.~37, p.~013002, Dec 2019.

\bibitem{Bahr:2009qc}
B.~Bahr and B.~Dittrich, ``{Improved and Perfect Actions in Discrete
  Gravity},'' {\em Phys. Rev.}, vol.~D80, p.~124030, 2009.

\bibitem{Dittrich:2014ala}
B.~Dittrich, ``{The continuum limit of loop quantum gravity - a framework for
  solving the theory},'' 2014.

\bibitem{Bahr:2014qza}
B.~Bahr, ``{On background-independent renormalization of spin foam models},''
  {\em Class. Quant. Grav.}, vol.~34, no.~7, p.~075001, 2017.

\bibitem{Ste:20}
S.~Steinhaus, ``Coarse graining spin foam quantum gravity -- a review,'' 2020.

\bibitem{KR86}
R.~Kadison and J.~Ringrose, {\em Fundamentals of the Theory of Operator
  Algebras. Volume II}.
\newblock Fundamentals of the Theory of Operator Algebras, American
  Mathematical Society, 1997.

\bibitem{Jan88}
J.~Janas, ``Inductive limit of operators and its applications,'' {\em Studia
  Mathematica}, vol.~90, pp.~87--102, 1988.

\bibitem{Sau98}
M.~L. Saunders, {\em {Categories for the Working Mathematician}}.
\newblock Graduate Texts in Mathematics, {\bf 5} (2nd ed.), Springer-Verlag New
  York, 1998.

\bibitem{Thi07}
T.~Thiemann, {\em Modern Canonical Quantum General Relativity}.
\newblock Cambridge Monographs on Mathematical Physics, Cambridge University
  Press, 2008.

\bibitem{Lang:2017beo}
T.~Lang, K.~Liegener, and T.~Thiemann, ``{Hamiltonian Renormalisation I:
  Derivation from Osterwalder-Schrader Reconstruction},'' 2017.

\bibitem{Dittrich:2011zh}
B.~Dittrich, F.~C. Eckert, and M.~Martin-Benito, ``{Coarse graining methods for
  spin net and spin foam models},'' {\em New J. Phys.}, vol.~14, p.~035008,
  2012.

\bibitem{Bahr:2011aa}
B.~Bahr, ``{Operator Spin Foams: holonomy formulation and coarse graining},''
  {\em J. Phys. Conf. Ser.}, vol.~360, p.~012042, 2012.

\bibitem{Bahr:2012qj}
B.~Bahr, B.~Dittrich, F.~Hellmann, and W.~Kaminski, ``{Holonomy Spin Foam
  Models: Definition and Coarse Graining},'' {\em Phys. Rev.}, vol.~D87, no.~4,
  p.~044048, 2013.

\bibitem{Dittrich:2012jq}
B.~Dittrich, ``{From the discrete to the continuous: Towards a cylindrically
  consistent dynamics},'' {\em New J. Phys.}, vol.~14, p.~123004, 2012.

\bibitem{Dittrich:2016tys}
B.~Dittrich, E.~Schnetter, C.~J. Seth, and S.~Steinhaus, ``{Coarse graining
  flow of spin foam intertwiners},'' {\em Phys. Rev.}, vol.~D94, no.~12,
  p.~124050, 2016.

\bibitem{Bahr:2016hwc}
B.~Bahr and S.~Steinhaus, ``{Numerical evidence for a phase transition in 4d
  spin foam quantum gravity},'' {\em Phys. Rev. Lett.}, vol.~117, no.~14,
  p.~141302, 2016.

\bibitem{Bahr:2017klw}
B.~Bahr and S.~Steinhaus, ``{Hypercuboidal renormalization in spin foam quantum
  gravity},'' {\em Phys. Rev.}, vol.~D95, no.~12, p.~126006, 2017.

\bibitem{Lang:2017yxi}
T.~Lang, K.~Liegener, and T.~Thiemann, ``{Hamiltonian Renormalisation II.
  Renormalisation Flow of 1+1 dimensional free scalar fields: Derivation},''
  2017.

\bibitem{Lang:2017oed}
T.~Lang, K.~Liegener, and T.~Thiemann, ``{Hamiltonian Renormalization III.
  Renormalisation Flow of 1+1 dimensional free scalar fields: Properties},''
  2017.

\bibitem{Lang:2017xrb}
T.~Lang, K.~Liegener, and T.~Thiemann, ``{Hamiltonian Renormalisation IV.
  Renormalisation Flow of D+1 dimensional free scalar fields and Rotation
  Invariance},'' 2017.

\bibitem{Liegener:2020dbc}
K.~Liegener and T.~Thiemann, ``{Hamiltonian Renormalisation V: Free Vector
  Bosons},'' 2020.

\bibitem{Thiemann:2020cuq}
T.~Thiemann, ``{Canonical Quantum Gravity, Constructive QFT and
  Renormalisation},'' 3 2020.

\bibitem{Morinelli:2020uea}
V.~Morinelli, G.~Morsella, A.~Stottmeister, and Y.~Tanimoto, ``{Scaling limits
  of lattice quantum fields by wavelets},'' 10 2020.

\bibitem{levin}
M.~Levin and C.~P. Nave, ``Tensor renormalization group approach to 2d
  classical lattice models,'' {\em Phys. Rev. Lett.}, vol.~99, p.~120601, 2007.

\bibitem{guwen}
Z.-C. Gu and X.-G. Wen, ``Tensor-{E}ntanglement-{F}iltering {R}enormalization
  {A}pproach and {S}ymmetry {P}rotected {T}opological {O}rder,'' {\em Phys.
  Rev. B}, vol.~80, p.~155131, 2009.

\bibitem{vidal-evenbly}
G.~Evenbly and G.~Vidal, ``{Tensor Network Renormalization},'' {\em Phys. Rev.
  Lett.}, vol.~115, p.~180405, 2015.

\bibitem{Dittrich:2013xwa}
B.~Dittrich and S.~Steinhaus, ``{Time evolution as refining, coarse graining
  and entangling},'' {\em New J. Phys.}, vol.~16, p.~123041, 2014.

\bibitem{Bahr:2009ku}
B.~Bahr and B.~Dittrich, ``{(Broken) Gauge Symmetries and Constraints in Regge
  Calculus},'' {\em Class. Quant. Grav.}, vol.~26, p.~225011, 2009.

\bibitem{friedlander1998introduction}
F.~Friedlander, G.~Friedlander, M.~Joshi, M.~Joshi, and M.~Joshi, {\em
  Introduction to the Theory of Distributions}.
\newblock Cambridge University Press, 1998.

\bibitem{reed1975ii}
M.~Reed and B.~Simon, {\em II: Fourier Analysis, Self-Adjointness}.
\newblock Methods of Modern Mathematical Physics, Elsevier Science, 1975.

\bibitem{Corichi:2007tf}
A.~Corichi, T.~Vukasinac, and J.~A. Zapata, ``{Polymer Quantum Mechanics and
  its Continuum Limit},'' {\em Phys. Rev. D}, vol.~76, p.~044016, 2007.

\end{thebibliography}
\bibliographystyle{ieeetr}
\end{document}